\documentclass[aps,pra,twocolumn,floatfix,superscriptaddress,nofootinbib,notitlepage,longbibliography]{revtex4-2}
\usepackage{amsmath,amssymb}
\usepackage{graphicx}
\usepackage{booktabs}
\usepackage{xcolor}
\usepackage[colorlinks=true,allcolors=blue]{hyperref}
\usepackage{siunitx}
\expandafter\def\csname ver@pdflscape.sty\endcsname{}
\usepackage{pdfpages}
\graphicspath{{figs/}}

\newcommand{\Nqubits}{8}
\newcommand{\Ncoup}{13}
\newcommand{\Nres}{200}
\newcommand{\Nstream}{6}
\newcommand{\Nstreamscal}{4}
\newcommand{\Ncue}{4}

\newcommand{\NSonset}{10^{4}}

\newcommand{\Jcircle}{0.32}
\newcommand{\Scircle}{64}
\newcommand{\Shcircle}{32}
 
\newcommand{\Nsurvey}{25}

\newcommand{\Ndamp}{36}
\newcommand{\Sclosed}{77}

\newcommand{\SdephMax}{77}
\newcommand{\gdephMax}{0.1}
\newcommand{\dSdeph}{+0.7}
\newcommand{\dSdephse}{0.4}
\newcommand{\SdephOne}{63}
\newcommand{\EdephClosed}{0.43}
\newcommand{\EdephPeak}{0.29}
\newcommand{\CdephClosed}{10.0}
\newcommand{\CdephPeak}{8.0}

\newcommand{\directbase}{50.7}
\newcommand{\directscal}{51.4}

\newcommand{\exactSQ}{97}

\newcommand{\ESNacueA}{42.1}

\newcommand{\ESNacuemax}{43.4}

\newcommand{\Nscalmin}{4}
\newcommand{\Nscalmax}{8}
\newcommand{\Nmaster}{100}

\newcommand{\SexactScalLo}{76}
\newcommand{\SexactScalHi}{97}

\newcommand{\SfiniteScalPeak}{78}
\newcommand{\SfiniteScalEight}{62}

\newcommand{\SfiniteScalKFour}{62}

\newcommand{\SfiniteScalKEight}{30}

 \newcommand{\Ecut}{E_N}

\begin{document}

\title{Quantum Reservoir Computing for Attention-like Selective Reproduction}

\author{Hadi Zadeh-Haghighi}
\email{hadi.zadehhaghighi@ucalgary.ca}
\affiliation{Department of Physics and Astronomy, University of Calgary, Calgary, Alberta, Canada}
\affiliation{Institute for Quantum Science and Technology, University of Calgary, Calgary, Canada}
\affiliation{Hotchkiss Brain Institute, University of Calgary, Calgary,  Canada}

\author{Christoph Simon}
\affiliation{Department of Physics and Astronomy, University of Calgary, Calgary, Alberta, Canada}
\affiliation{Institute for Quantum Science and Technology, University of Calgary, Calgary, Canada}
\affiliation{Hotchkiss Brain Institute, University of Calgary, Calgary,  Canada}

\begin{abstract}
\noindent
Choosing which information to use as the context changes is a basic
cognitive operation. We introduce a new task of this kind for quantum
reservoir computing (QRC): a binary cue indicates which of two
simultaneous input streams the reservoir must reproduce. In simulated
spin reservoirs with a trained linear readout, infinitely many
measurements (exact expectation values) let the reservoir reproduce
the cued stream accurately. At a finite number of measurements
\(N_M\), the score exceeds that of a linear readout of the raw inputs
only at large \(N_M\) because of sampling noise.
Performance increases with both entanglement and coherence in the
computational basis, and then saturates. These results show that
QRC can select and reproduce a cued stream, and they are a starting
point for harder cognitive tasks.

\end{abstract}

\maketitle

\section{Introduction}

Selecting relevant information according to context is a basic component
of cognition. The same sensory input can be relevant in one context and
distracting in another, so successful processing requires combining
information about the input with an instruction about what to report.
Following a particular speaker among competing voices provides a
familiar motivation for such selective
processing~\cite{osullivan2015attentional}. Context-dependent
computation has been studied in cortical dynamics and in trained
recurrent neural
networks~\cite{mante2013context,song2016training}. Classical reservoir
computing has been applied to related cognitive operations: a
prefrontal reservoir model uses an explicit context
signal~\cite{enel2016reservoir}, and connectome-based reservoirs have
been tested on NeuroGym context-dependent decision making, in which a
cue selects which of two perceptual discriminations to
perform~\cite{suarez2024conn2res}.
Reservoir computing offers a natural setting for context-dependent
selection because the cue has to be combined with the inputs
nonlinearly: a fixed dynamical system can form that combination in its
features, and only a simple readout is
trained~\cite{jaeger2004harnessing,maass2002real,lukovsevivcius2009reservoir}.

Quantum reservoir computing (QRC) implements this transformation through
the dynamics of a driven quantum
system~\cite{fujii2017harnessing,mujal2021opportunities}, with
realizations including spin networks, nonlinear oscillators, Rydberg
arrays, NMR, and circuit
QED~\cite{martinezpena2021dynamical,govia2021,bravo2022,liu2026rydberg,hou2026,markovic2026experimental}.
There is now a substantial literature on its memory, dynamical
regimes, and physical
implementation~\cite{tanaka2019recent,martinezpena2023quantum,kobayashi2026edge,tariq2026review}.
Much of that work has centered on conventional temporal
tasks: short-term memory, NARMA-type recurrences, and chaotic
forecasting~\cite{fujii2017harnessing,nakajima2019boosting,pfeffer2022hybrid,ahmed2024prediction,ricci2026damping}.
Rydberg QRC has also been used for cognitive-style operations such as
decision and working memory in a trained recurrent
setting~\cite{bravo2022}.

In other RC work, ``attention'' refers to an architecture rather than
to context-dependent selection: an attention layer on the readout for
chaotic prediction~\cite{koster2024attention}, or trained gates that
retain or discard inputs~\cite{sakemi2024selfmodulated}. Those
architectures, and the QRC studies above, motivate a concrete question:
can a \emph{fixed} quantum dynamical system, with only a linear readout
trained, carry out that selection?

Here we introduce a new task for QRC, which we call
\emph{attention-like selective reproduction}. Context-dependent selection is already
studied in cortex and in trained recurrent
networks~\cite{mante2013context,song2016training}, and related reservoir
tasks ask for a trial-based
choice~\cite{enel2016reservoir,suarez2024conn2res}. The new part is
reproduction of the selected stream. Two matched streams arrive
together, a binary cue indicates which stream to reproduce, and the
output at every step is the current value of the cued stream.

We implement the task numerically in small disordered spin reservoirs.
The inputs enter through one qubit, the spins evolve under a fixed
Ising Hamiltonian, and a trained linear readout is built from measurements
of the spins. A setting is successful when the score exceeds a linear readout of
the raw inputs and the valid-cue score exceeds the shuffled-cue score.

Previous QRC studies have associated computational performance with
entanglement, coherence, the explored state space, non-Markovianity,
and other quantumness measures, while also showing that the predictive
value of those measures depends on the task, input timescale, dynamical
regime, dissipation, and finite
sampling~\cite{gotting2023exploring,kora2024frequency,palacios2024role,karimi2026role,kora2026noise,sannia2024,kobayashi2024influx,garciabeni2024,gotting2025,sannia2026nonmarkov,cindrak2025krylov,xia2026magic,ivaki2025,askari2025distributed}.
A related spin-network task injects two streams on separate qubits and
targets their product, with no selecting
cue~\cite{askari2025distributed}. We quantify entanglement by the
logarithmic negativity of a predeclared three-qubit input block versus
the remaining qubits~\cite{vidal2002computable,plenio2005logarithmic},
and coherence by the computational-basis \(\ell_1\)
measure~\cite{baumgratz2014}.

With exact readout the reservoir reproduces the cued stream. At a
finite \(N_M\) it does so only when enough measurements are
available~\cite{hu2023tackling,palacios2024role,kora2026noise,liu2026rydberg},
and a larger reservoir can then score lower because of sampling noise.
Performance increases with entanglement and coherence, but under
dephasing the score can remain above a linear readout of the raw
inputs after both markers have vanished at the readout. The sections
below describe the new task in detail, and give the dependence on the
number of measurements, the reservoir size, and dephasing.

\section{Results}
\label{sec:results}

\subsection{Attention-like selective reproduction with a quantum reservoir}
\label{sec:taskdef}
\label{sec:atlas}

\begin{figure*}[t]
  \centering
  \includegraphics[width=0.92\textwidth]{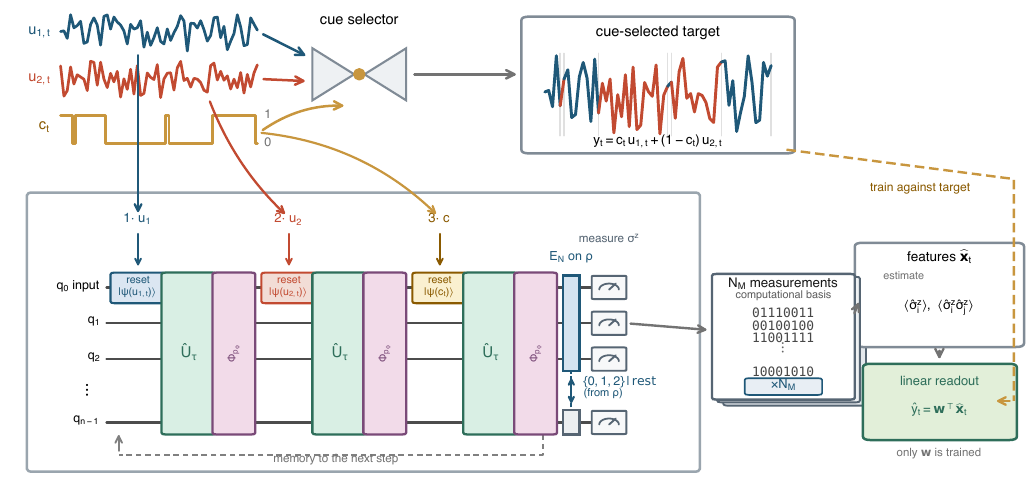}
  \caption{\textbf{Using a quantum reservoir for context-dependent selective reproduction.}
    The two streams and the cue in Eq.~\eqref{eq:target} are
    sequentially encoded through reset operations on qubit \(q_0\).
    Each reset is followed by the propagator
    \(\hat U_\tau=e^{-i\hat H\tau}\) of the Hamiltonian in
    Eq.~\eqref{eq:H} and by a local phase flip of probability
    \(p_{\varphi}\) on every qubit
    (the identity at \(\gamma_\varphi=0\); see Methods, Sec.~\ref{sec:local-deph}). In the Supplement,
    relaxation toward \(\lvert 0\rangle\) occupies the same slot as
    this dephasing. A record of \(N_M\) computational-basis measurements
    supplies all one- and two-body \(\sigma^z\) features. Only the
    linear readout is trained. At the step boundary, before those
    measurements, the main-text entanglement diagnostic is the
    logarithmic negativity of a predeclared three-qubit input block
    \(\{0,1,2\}\) versus the remaining qubits: qubit \(0\) is the
    injection site, and qubits \(1,2\) complete that block because all
    three inputs enter through \(q_0\). Other partitions are in the
    Supplement. Entanglement is computed from the simulated state, not
    from the measurement record.}
  \label{fig:method}
\end{figure*}

Let \(u_{1,t}\) and \(u_{2,t}\) be two independent streams of i.i.d.\ uniform
noise, each standardized to zero mean and unit variance, and let
\(c_t\in\{0,1\}\) be a binary cue. The target is
\begin{equation}
  y_t = c_t\,u_{1,t} + (1-c_t)\,u_{2,t}.
  \label{eq:target}
\end{equation}
Neither stream is the signal on its own: the cue assigns which one
to reproduce.
We implement the task numerically in disordered spin reservoirs with
\(n=\Nscalmin\)--\(\Nscalmax\) qubits (Fig.~\ref{fig:method}). The
reservoir Hamiltonian is the transverse-field XX Ising model
\begin{equation}
\hat H=\sum_{i<j} J_{ij}\,\hat\sigma^x_i\hat\sigma^x_j + h\sum_i \hat\sigma^z_i.
\label{eq:H}
\end{equation}
Within each logical step the two stream values and
the cue are encoded into one input qubit, and between injections the
spins evolve under the propagator \(\hat U_\tau=e^{-i\hat H\tau}\) of
that Hamiltonian. The readout is a trained linear combination of one-
and two-body \(\sigma^z\) observables. Methods specifies the spin
architecture and the measurement protocol for that reservoir.

The cue is present at every step, so Eq.~\eqref{eq:target} is the
current value of the cued stream. Rewriting it separates a fixed mix
of the two streams, \((u_{1,t}+u_{2,t})/2\), from the product
\((c_t-\tfrac12)(u_{1,t}-u_{2,t})\).

The score on the held-out test window is \(S=100R^2\), where \(R\) is the
Pearson correlation between the readout prediction \(\tilde y_t\) and the
target \(y_t\). A perfect reconstruction scores \(100\). On the
eight-qubit reservoirs, a linear readout of the two streams and the
cue scores \(\directbase\) out of \(100\). A bilinear readout that is
also given the product of the cue and the streams reaches a perfect
score. A setting is successful when the reservoir
score exceeds the linear readout and the valid-cue score exceeds a
shuffled-cue control, which uses the same reservoir and streams with
the cue scrambled.

\begin{figure}[htbp]
  \centering
  \includegraphics[width=0.88\columnwidth]{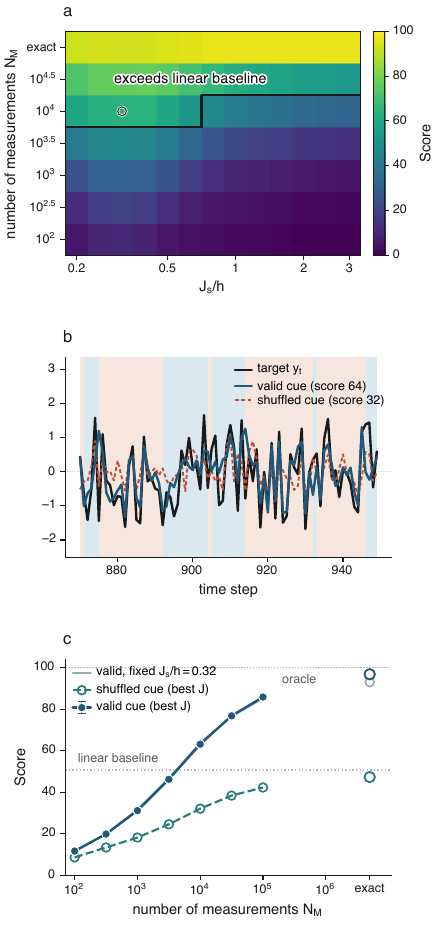}
  \caption{\textbf{Number of measurements controls selection performance}
    (\(n=8\) qubits, no added noise).
    \textbf{(a)}~Mean valid-cue score versus coupling \(J_s/h\) and
    number of measurements \(N_M\). The top row uses exact expectation
    values. Cells whose mean score exceeds a linear readout of the raw
    inputs (\(\directbase\)) are labeled in the plot; the outline follows
    the edges of those cells on the measured grid. The open circle is
    \(J_s/h=\Jcircle\), \(N_M=10^4\), used in (b) and (c).
    \textbf{(b)}~Test traces at that point. Background: the cue selects
    \(u_1\) (blue) or \(u_2\) (red).
    \textbf{(c)}~Score versus \(N_M\). Filled circles: at each
    \(N_M\), coupling is chosen to maximize validation score. Grey:
    coupling held at the circle in (a). Open circles: exact readout.
    The two open circles at the top are the valid-cue scores at the
    best \(J\) and at the fixed \(J\) of (a). The connected point at
    \(N_M=10^5\) is drawn to continue the curve. It is not used to
    choose the coupling or the outline in (a).}
  \label{fig:atlas}
\end{figure}

With infinitely many measurements (exact expectation values), the
eight-qubit reservoir reaches a valid-cue score of \(\exactSQ\) over a
wide range of coupling (Fig.~\ref{fig:atlas}a). The measured features
therefore contain the nonlinear selection rule, even though the
readout itself is linear. At small \(N_M\), sampling noise hides that
information and the score lies below a linear readout of the raw
inputs. At the marked coupling \(J_s/h=\Jcircle\) and \(N_M=10^4\),
the valid-cue output follows the selected stream, with mean scores
\(\Scircle\) (valid cue) and \(\Shcircle\) (shuffled cue)
(Fig.~\ref{fig:atlas}b). The score rises with \(N_M\) and crosses the
linear readout near \(N_M=\NSonset\) (Fig.~\ref{fig:atlas}c).

Reading the reservoir at several times within a step does not remove
this sampling limit when the total number of measurements is held
fixed (Supplementary Fig.~S4).

\subsection{Effects of reservoir size and number of measurements}
\label{sec:scaling}

\begin{figure}
  \centering
  \includegraphics[width=\columnwidth]{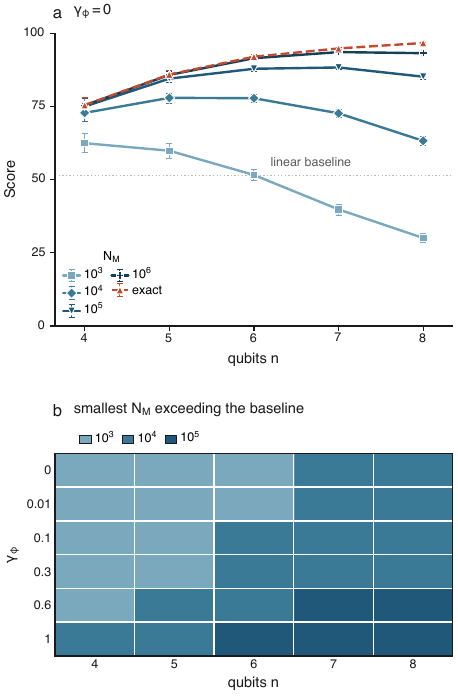}
  \caption{\textbf{Reservoir size and number of measurements jointly determine
    accessible performance.}
    Ensemble means are evaluated at \(J^\star(n,N_M)\), the coupling
    that maximizes mean score at \(\gamma_\varphi=0\).
    Each random eight-qubit coupling matrix is reused at
    smaller \(n\) by keeping only the first \(n\) qubits, so the sizes
    share the same couplings and the same input qubit.
    \textbf{(a)}~Score versus \(n\) at \(\gamma_\varphi=0\), for
    \(N_M=10^3\)--\(10^6\) and exact readout. Error bars are \(95\%\)
    intervals from resampling reservoirs. Dotted line:
    linear baseline \(\directscal\).
    \textbf{(b)}~Smallest tested \(N_M\) at which the mean score
    exceeds that baseline as \(\gamma_\varphi\) increases.
    The coupling chosen at \(\gamma_\varphi=0\) is held fixed; it is
    not re-chosen under dephasing. The discrete colors denote \(10^3\),
    \(10^4\), and \(10^5\). None of the displayed conditions first
    crosses the baseline at \(N_M=10^6\). Scores versus size at
    nonzero dephasing are shown in Supplementary Fig.~S9.}
  \label{fig:sizesum}
\end{figure}

In Fig.~\ref{fig:sizesum} the task is repeated for
\(n=\Nscalmin\)--\(\Nscalmax\) qubits. Each
random eight-qubit coupling matrix is reused at smaller sizes by
keeping only the first \(n\) qubits, so the sizes share the same
couplings and the same input qubit. For each pair \((n,N_M)\),
\(J^\star(n,N_M)\) is the coupling that maximizes the mean score at
\(\gamma_\varphi=0\). That coupling is then held fixed as dephasing
increases; it is not chosen again. This survey uses a different set of
inputs. A linear readout of those raw inputs scores \(\directscal\).
The survey includes all \(\Nsurvey\) predefined combinations of size and
\(N_M\), including those whose scores at finite \(N_M\) fall below
that linear readout.

Without added dephasing, infinitely many measurements (exact
expectation values) improve the score from
\(\SexactScalLo\) at \(n=4\) to \(\SexactScalHi\) at \(n=8\)
(Fig.~\ref{fig:sizesum}a). A finite \(N_M\) reverses
that trend. At \(N_M=10^3\), the score falls from
\(\SfiniteScalKFour\) to \(\SfiniteScalKEight\) as \(n\) grows. At
\(N_M=10^4\), every size exceeds a linear readout of the raw inputs,
but the score peaks near \(\SfiniteScalPeak\) at \(n=5\) and falls to
\(\SfiniteScalEight\) at \(n=8\). A larger set of exact features
therefore helps only if enough measurements are available to read them.

Dephasing generally increases the \(N_M\) needed to
exceed a linear readout of the raw inputs (Fig.~\ref{fig:sizesum}b).
At \(\gamma_\varphi=0.6\), four qubits still do so at \(N_M=10^3\),
while seven and eight qubits need \(N_M=10^5\) on the grid we tested.
At complete dephasing, \(\gamma_\varphi=1\), every size is below that
line at \(N_M=10^3\). The smallest successful value on the grid is then
\(10^4\) for \(n=4,5\) and \(10^5\) for \(n=6\)--\(8\). These are the
smallest values on the grid at which the score exceeds the baseline,
not a continuous estimate of that threshold.

Exact readout stays accurate even at complete dephasing
(Supplementary Fig.~S9). With exact readout the reservoir still
selects the cued stream when dephasing is applied after each injection.
The score no longer rises steadily with size. Holding the number of
readout features fixed removes most of the exact-readout gain with
size, so much of that gain comes from measuring more observables
(Supplementary Fig.~S3). A classical echo-state network with a
comparable number of features also performs the task
(Supplementary Fig.~S2).

\subsection{Performance under dephasing}
\label{sec:marker}

\begin{figure}
  \centering
  \includegraphics[width=\columnwidth]{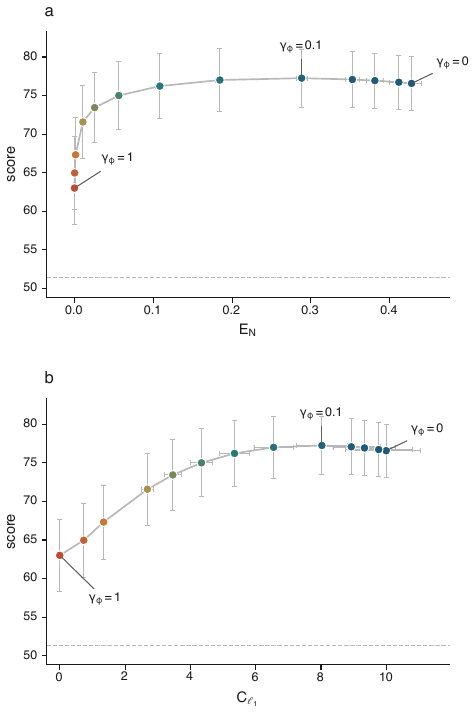}
  \caption{\textbf{Score increases with entanglement and coherence, then
    saturates.}
    Each point is the equal-weight mean over the \(\Nsurvey\)
    combinations of reservoir size (\(n=4\)--\(8\)) and number of
    measurements (\(N_M=10^{3}\)--\(10^{6}\) and exact readout).
    Coupling is \(J^\star(n,N_M)\), chosen with no added dephasing;
    the same reservoirs are then followed as dephasing increases.
    \textbf{(a)}~Score versus logarithmic negativity \(E_N\) of the
    predeclared input block \(\{0,1,2\}\) versus the remaining qubits
    (qubit \(0\) is the injection site; all three inputs enter through
    it). A two-qubit singlet has \(E_N=1\). For this cut, \(E_N\) is at
    most \(\min(3,n-3)\) ebits, the number of qubits on the smaller
    side. Other three-qubit cuts on the mixed-stream atlas, rather
    than this dephasing sweep, are in Supplementary Fig.~S5.
    \textbf{(b)}~Score versus computational-basis \(\ell_1\) coherence
    \(C_{\ell_1}\) (\(0\) if the state is diagonal). Matching points in
    (a) and (b) share the same dephasing and the same score. Error
    bars show variation among the \(\Nsurvey\) combinations. Labels
    mark \(\gamma_\varphi=0\), \(\gamma_\varphi=\gdephMax\), and
    complete dephasing. Both are associated with a higher score.
    The rise with entanglement levels off while score is still
    changing with coherence. At complete dephasing both quantities
    vanish at readout and the mean score remains \(\SdephOne\), above
    the linear baseline \(\directscal\).}
  \label{fig:deph}
\end{figure}

In Fig.~\ref{fig:deph} we ask whether entanglement and coherence
distinguish reservoirs that have the same size, the same number of
measurements, and the same coupling, and how those links change under
dephasing. The survey has
\(\Nsurvey\) conditions: \(n=4\)--\(8\) at
\(N_M\in\{10^3,10^4,10^5,10^6,\infty\}\). A supplementary regression
of entanglement against performance uses only the eight-qubit settings
that pass the success test of Sec.~\ref{sec:atlas}. Within each condition the
coupling stays at \(J^\star(n,N_M)\) while dephasing increases.
Entanglement is the logarithmic negativity \(E_N\) of qubits
\(\{0,1,2\}\) versus the rest. Coherence is the computational-basis
\(\ell_1\) sum \(C_{\ell_1}\). Both are computed from the state before
readout (see Methods, Secs.~\ref{sec:ent-def} and~\ref{sec:coh-def}). For a fixed trajectory, neither depends on the
measurement record or on \(N_M\). Relaxation is off in this figure.
Supplementary Fig.~S8 instead applies relaxation toward
\(\lvert 0\rangle\), with dephasing off. Other three-qubit cuts, from the mixed-stream
atlas rather than along this dephasing sweep, are in
Supplementary Fig.~S5.

At \(\gamma_\varphi=0\), the mean score across the \(\Nsurvey\)
conditions is \(\Sclosed\), with mean \(E_N=\EdephClosed\) and mean
\(C_{\ell_1}=\CdephClosed\) (Fig.~\ref{fig:deph}). Among reservoirs in the same condition,
higher entanglement is associated with a higher score. Mild
dephasing lowers entanglement and coherence but barely changes the
mean score. At \(\gamma_\varphi=\gdephMax\), mean \(E_N\) falls to
\(\EdephPeak\) and mean \(C_{\ell_1}\) to \(\CdephPeak\), while the
mean score stays near \(\SdephMax\). The change from
\(\gamma_\varphi=0\) is \(\dSdeph\pm\dSdephse\) points. Reservoirs in the
same condition still tend to score higher when their entanglement is
higher.

At complete dephasing, both quantities are consistent with zero at
readout, yet the mean score remains \(\SdephOne\), above the linear
readout of the raw inputs (\(\directscal\); Fig.~\ref{fig:deph}).
Along the sweep the mean score rises with both, but the curves have
different shapes. Against entanglement the score levels off, and
\(E_N\) is already negligible while coherence and score are still
changing. Coherence remains appreciable over a wider range of
dephasing. Both curves come from this single dephasing sequence.
The two quantities also have different ranges, and the
coherence sample uses fewer reservoir seeds, so the comparison of
shapes is qualitative.

\section{Discussion}
\label{sec:discussion}

The task asks a quantum reservoir to do something familiar from
cognition: use a cue to decide what is relevant, then reproduce that
stream with one linear readout. With infinitely many measurements
(exact expectation values), a small spin
reservoir can form the needed product of the cue and the streams.
The task is one member of an established family of context-dependent
computations~\cite{mante2013context,song2016training,suarez2024conn2res}, and a concrete place to start on harder
cognitive tasks in QRC. Related classical reservoir tasks are
trial-based binary choices, not continuous reproduction of a cued
stream~\cite{enel2016reservoir,suarez2024conn2res}.

A central distinction is between what the reservoir represents and
what a finite \(N_M\) can recover. Sampling noise is
a known obstacle to reading QRC
features~\cite{hu2023tackling,palacios2024role,kora2026noise}.
Across the sizes we studied, larger \(n\) improves the
expectation-value score, but a larger reservoir can score lower at
small \(N_M\).
Dephasing raises the number of measurements needed for a useful score
(Fig.~\ref{fig:sizesum}). Two controls support that interpretation. Much of
the gain with size disappears when the number of readout features is
held fixed (Supplementary Fig.~S3). That check mixes the two streams
before they are injected, with weight \(\alpha=0.3\).
The coefficient \(\alpha\) is defined in Supplementary Eq.~(S1);
\(\alpha=0\) leaves the streams unmixed, as in the main figures.
Reading the state at extra times does
not remove the sampling limit when the total number of measurements is
fixed (Supplementary Fig.~S4). Size, which observables are used, and
how many times the register is measured therefore have to be
considered together.

Entanglement and coherence are linked to the score, but only in a
qualified way. Without added dephasing, and at the same size, number
of measurements, and coupling, higher entanglement is modestly
correlated with a higher score. Along the dephasing sweep, the mean
score rises with both entanglement and coherence, and it levels off
sooner against entanglement (Fig.~\ref{fig:deph}). The two quantities
are two descriptions of the same sequence of states.

At complete dephasing the state is diagonal in the computational basis,
so both quantities vanish at readout, yet the mean score remains
\(\SdephOne\), above a linear readout of the raw inputs. After each
reset, and before dephasing, the
register evolves under the spin Hamiltonian. That interval can
generate coherence and entanglement, and it can change which
computational-basis states are occupied. Dephasing then removes the
coherences, the off-diagonal elements, and the readout sees only the
probabilities of those basis states. The score is computed from those
probabilities. That association also weakens near a
perfect score, and part of it is present even with a shuffled cue
(Supplementary Figs.~S6 and~S7).

Supplementary Fig.~S8 replaces dephasing with relaxation toward
\(\lvert 0\rangle\). Moderate relaxation can raise the mean score
while lowering \(E_N\); strong relaxation erases the state and drives
both to zero. The intermediate improvement might come from a shorter
memory, from fewer poorly resolved features, or from a more favorable
dynamical regime. Dissipation
has been studied as a computational ingredient in other QRC
work~\cite{sannia2024,ricci2026damping,gotting2025,sannia2026nonmarkov}.

Each measurement is modeled
by restarting the reservoir and replaying the inputs up to that time.
For \(N_M=10^4\) and \(T=1200\) steps, that is \(1.2\times10^7\)
measured runs and about \(7.2\times10^9\) logical-step executions
(see Methods, Sec.~\ref{sec:finite-nm}). Errors in state preparation,
reset, gates, and measurement, together with drift, are left out.
Those errors have to be assessed for a particular device.

A bilinear model and an echo-state network show what the task demands.
The bilinear model, given the product of the cue and the
streams, solves the target exactly. A \(36\)-node echo-state network,
matching the \(36\)
quantum features at \(n=8\), also does the task, with a valid-cue gain
of \(\ESNacueA\)--\(\ESNacuemax\) points (Supplementary Fig.~S2a). Both the
echo-state network and the quantum curves in that figure are read exactly.
In the main figures the quantum features are estimated from measurements. For the reservoirs studied here, the
classical network is much cheaper: it is an ordinary update, while
each quantum step is estimated by restarting and remeasuring the
register. A quantum advantage is not expected at this size, on this
task. It becomes plausible only for much larger quantum reservoirs and
for tasks much harder than selecting one of two streams.

The demonstration is numerical, uses synthetic streams, and covers
\(n=\Nscalmin\)--\(\Nscalmax\) qubits. The readout uses only one- and
two-body \(\sigma^z\) observables. The main figures report entanglement
for one split, qubits \(\{0,1,2\}\) versus the rest. Other three-qubit
cuts and the one-qubit input cut are in Supplementary Figs.~S5 and~S7,
on mixed-stream scans rather than along the dephasing sweep. Coherence
is reported in the computational basis only. All three
inputs enter one qubit in a fixed order. Swapping the two stream ports
leaves the score unchanged. Reversing the injection order is nearly
equivalent at exact readout and improves the score when the number of
measurements is finite. Moving the stream injections closer to the
measurement is a possible explanation for that improvement, and the
mechanism remains open (Supplementary Table~S1). In the
dephasing and relaxation surveys, the coupling is chosen without noise
and then held fixed. Adapting the task to hardware therefore means
choosing the input encoding and the measurement access together.

A fixed quantum evolution, with only a linear readout trained, can form
a contextual selection. The physical dynamics mix the cue with the
inputs, and only the readout is trained. That is a neuromorphic use of a
quantum reservoir. A finite measurement record then has to recover that
selection, and a larger register can make the recovery harder. The
broader target is to implement more cognitive tasks on quantum
neuromorphic architectures of this kind. Decision making and working
memory have already been studied with trained recurrent quantum
reservoirs, and context-dependent choice is a standard task for
classical
networks~\cite{bravo2022,mante2013context,song2016training}. The next
step is to implement that family with a fixed evolution on a spin,
Rydberg, or circuit device, and to keep the outcome visible in a
finite measurement record. Beyond that family are computations in
which the device forms a context from its own recent input and puts
that context to further use, with the same requirement that the result
remain visible once the state is read only a finite number of times.
That is one path toward quantum artificial intelligence: the quantum
evolution supplies the computation, and a trained readout makes it
available.

\section{Methods}
\label{sec:methods}

\subsection{Reservoir computing and the spin-QRC architecture}
\label{sec:architecture}

This section specifies the reservoir and how it is measured. The
performance maps and the entanglement analysis in the main text use
this setup. Figure~\ref{fig:method} summarizes it: a fixed dynamical
system turns the inputs into a large set of features, and only a
linear readout is trained.

\subsubsection{The reservoir-computing pipeline}
A reservoir computer maps an input sequence \(u_t\) to an output \(\tilde y_t\) through
\begin{equation}
u_t \;\longrightarrow\; \text{reservoir} \;\longrightarrow\;
\mathbf{x}_t \;\longrightarrow\; \tilde y_t=\mathbf{w}^\top\mathbf{x}_t.
\end{equation}
The reservoir is a fixed nonlinear dynamical system. It keeps a fading
memory of past inputs and expands the current input nonlinearly.
\(\mathbf{x}_t\) is the vector of features read at step \(t\), and
\(\mathbf{w}\) is a trained linear
readout~\cite{jaeger2004harnessing,maass2002real}. The reservoir
itself is never trained for the task. All learning is in the linear
map \(\mathbf{w}\), fit by ridge regression, with the penalty chosen
on a validation split.

\subsubsection{Spin reservoir and data injection}
The reservoir is an \(n\)-qubit disordered spin network. Throughout, hats denote quantum operators
(\(\hat H\), \(\hat U_\tau\), \(\hat\sigma^x_i\), \(\hat\sigma^z_i\), \(\hat\rho\)), calligraphic symbols denote
channels (superoperators), and scalar quantities (\(u_t\), \(y_t\), \(J_{ij}\), \(N_M\), scores)
carry no hat. A bar on \(\bar{x}_k\) is a sample average. A tilde on
\(\tilde y_t\) is the readout prediction. One logical step of the resulting
circuit is drawn in Fig.~\ref{fig:method}. The
internal dynamics are generated by the transverse-field XX Ising
Hamiltonian of Eq.~\eqref{eq:H},
with all-to-all couplings \(J_{ij}\) drawn once per reservoir realization and held fixed. We report
the coupling strength through the dimensionless ratio \(J_s/h\); \(h\) sets the energy scale and
\(\tau\) the evolution time per substep. Between injections the state evolves under
\(\hat U_\tau=e^{-i\hat H\tau}\), i.e.\ under the unitary channel
\(\mathcal{U}_\tau(\hat\rho)=\hat U_\tau\hat\rho\,\hat U_\tau^{\dagger}\).

Data enter the quantum state by \emph{reset injection} on the input qubit (qubit \(0\)): the reset
channel replaces that qubit with a pure state whose amplitude encodes the scalar input while the
remaining qubits carry the reservoir memory forward,
\begin{align}
\mathcal{R}_x^{(0)}(\hat\rho)&=\lvert\psi(x)\rangle\langle\psi(x)\rvert_0\otimes
\operatorname{Tr}_0(\hat\rho),
\nonumber\\
\lvert\psi(x)\rangle&=\sqrt{1-f(x)}\,\lvert0\rangle+\sqrt{f(x)}\,\lvert1\rangle,
\label{eq:reset}
\end{align}
where \(f\) places the signal between 0 and 1. For a stream,
\(0.25u+0.5\) is used when it lies in that interval. A value below 0
is set to 0, and a value above 1 is set to 1.
The cue is already 0 or 1, so \(f(c)=c\). The three task
signals of one logical step---the two streams \(u_1,u_2\) and the cue \(c\)---are
injected as three sequential resets, each followed by an evolution under
\(\mathcal{U}_\tau\) and, when dephasing is on, the local dephasing channel
\(\Phi_{p_{\varphi}}^{\otimes n}\) defined below
(Fig.~\ref{fig:method}; the identity when \(\gamma_\varphi=0\)). Each reset writes one value
into qubit~$0$, so the three values are injected one after another on
that qubit.

One logical step advances
the density operator through three substeps,
\begin{align}
\hat\rho_{t,0}&=\hat\rho_t^{-},
\nonumber\\
\hat\rho_{t,1}&=\Phi_{p_{\varphi}}^{\otimes n}\!\circ\mathcal{U}_\tau\!\circ
\mathcal{R}_{u_{1,t}}^{(0)}(\hat\rho_{t,0}),
\nonumber\\
\hat\rho_{t,2}&=\Phi_{p_{\varphi}}^{\otimes n}\!\circ\mathcal{U}_\tau\!\circ
\mathcal{R}_{u_{2,t}}^{(0)}(\hat\rho_{t,1}),
\nonumber\\
\hat\rho_{t,3}&=\Phi_{p_{\varphi}}^{\otimes n}\!\circ\mathcal{U}_\tau\!\circ
\mathcal{R}_{c_t}^{(0)}(\hat\rho_{t,2}),
\label{eq:statecycle}
\end{align}
with \(\hat\rho_{t+1}^{-}=\hat\rho_{t,3}\) (Fig.~\ref{fig:method}). At
\(p_{\varphi}=0\) the dephasing channel is the identity, so
evolution between injections is unitary; the reset operations remain
part of the driven, nonunitary input protocol.

\subsubsection{Local dephasing}
\label{sec:local-deph}
The main-text noise channel is local dephasing, acting after
\(\mathcal{U}_\tau\) following each injection. On each qubit it is a
phase flip with probability \(p_{\varphi}\),
\begin{equation}
\hat L_0=\sqrt{1-p_{\varphi}}\,\hat I,
\qquad
\hat L_1=\sqrt{p_{\varphi}}\,\hat\sigma^z,
\label{eq:kraus-deph}
\end{equation}
so
\(\Phi_{p_{\varphi}}^{(i)}(\hat\rho)
=(1-p_{\varphi})\hat\rho
+p_{\varphi}\,\hat\sigma^z_i\hat\rho\,\hat\sigma^z_i\).
The \(n\)-qubit map is the tensor product of the local channels. In the
computational basis that product multiplies every matrix element by a Hamming
factor,
\begin{equation}
\bigl[\Phi_{p_{\varphi}}^{\otimes n}(\hat\rho)\bigr]_{ab}
=(1-2p_{\varphi})^{d_H(a,b)}\rho_{ab},
\label{eq:deph-mask}
\end{equation}
with \(d_H(a,b)\) the Hamming distance of bitstrings \(a\) and \(b\).
Each application preserves diagonal populations and suppresses
off-diagonal elements. Complete dephasing
is \(p_{\varphi}=1/2\) (\(1-2p_{\varphi}=0\)), not
unity: \(p_{\varphi}=1\) is a global \(\hat\sigma^z\) and is
unitary. The channel follows each of the three injections, so in
one logical step a coherence of Hamming distance \(d_H\) is
multiplied by \((1-2p_{\varphi})^{d_H}\) three times. We report the
strength of that whole step on a \(\Ndamp\)-point ladder
\(\gamma_\varphi\in[0,1]\),
\begin{equation}
p_{\varphi}=\tfrac12\bigl[1-(1-\gamma_\varphi)^{1/3}\bigr],
\label{eq:gamma-phi}
\end{equation}
which sets \(1-2p_{\varphi}=(1-\gamma_\varphi)^{1/3}\). The cube
root gives each injection the same factor, and the product of the
three is \((1-\gamma_\varphi)^{d_H}\). A single-qubit coherence is
therefore multiplied by \(1-\gamma_\varphi\) over the step.
\(\gamma_\varphi=1\) removes it: a phase flip with probability
\(1/2\) after every injection.
Because input encoding and evolution can continue to alter the
populations, complete dephasing does not force the reservoir into
\(\lvert 0\cdots 0\rangle\) or eliminate its diagonal readout features.

\subsubsection{Readout with a finite number of measurements}
The readout features are the one- and two-body computational-basis correlators
\(\langle \hat\sigma^z_i\rangle\) and \(\langle \hat\sigma^z_i\hat\sigma^z_j\rangle\),
giving \(d(n)=n(n+1)/2\) observables
(\(d=10\) at \(n=4\), \(d=36\) at \(n=8\)). Exact expectation-value readout evaluates
\(x_k(t)=\operatorname{Tr}[\hat O_k\hat\rho_t]\) for each observable \(\hat O_k\). All of these
observables are diagonal in the \(\sigma^z\) basis and therefore commute, so a single projective
measurement of the register yields one computational-basis bitstring, and a set of \(N_M\) such
bitstrings estimates the entire feature vector simultaneously through the statistical estimator
\begin{equation}
\bar{x}_k(t)=\frac{1}{N_M}\sum_{s=1}^{N_M} m_k^{(s)}(t),
\end{equation}
with \(m_k^{(s)}\) the value of observable \(k\) on measurement \(s\).
The estimates carry correlated sampling noise, because every one-
and two-body feature is reconstructed from the same bitstrings; the exact-readout limit
\(N_M\to\infty\) recovers \(x_k\). Readout-time multiplexing samples one
logical step at several equally spaced times. Each sample is a virtual
node, and \(V\) is the number of samples. Every extra sample is a further
measurement of that step. The main-text performance map uses one sample
per step (\(V=1\)). The control with more than one sample (\(V>1\)) is
Supplementary Fig.~S4. We vary \(N_M\) together with \(J_s/h\) and \(n\).
Exact readout, written \(N_M=\infty\), is kept as its own reference and
is not placed on the finite \(N_M\) axis.

\subsection{Context-dependent selection task}
\label{sec:task_and_model}
\label{sec:task}

The attention-like selection task is defined in Sec.~\ref{sec:taskdef}.
Equation~\eqref{eq:target} is the cued stream.

\subsubsection{Reservoir settings}
\label{sec:reservoir}

The spin reservoir, its sequential substep injection of \((u_1,u_2,c)\) on the
input qubit, and its \(\langle \hat\sigma^z_i\rangle,\langle \hat\sigma^z_i\hat\sigma^z_j\rangle\)
readout are exactly as described in Sec.~\ref{sec:architecture}
[Eqs.~\eqref{eq:H}--\eqref{eq:statecycle}]. The \(n=\Nqubits\) atlas map
varies coupling and \(N_M\) at fixed size; the size survey uses nested
reservoirs \(n=\Nscalmin\)--\(\Nscalmax\). Main-text figures leave the two streams unmixed (\(\alpha=0\)).
In the supplementary controls each injected stream is a mixture of
both inputs, with strength \(\alpha\); the normalized definition is
Supplementary Eq.~(S1).
All other dynamical parameters
are held fixed across parameter settings. In the main text the only
quantities that vary are the coupling ratio \(J_s/h\) and the number of
measurements \(N_M\). Each logical step comprises
three reset sub-steps, each followed by evolution for a fixed duration
\(\Delta t = 3\). 

Main-text entanglement \(\Ecut\) is defined in
Sec.~\ref{sec:ent-def}. Because all three scalars \((u_1,u_2,c)\)
enter through qubit~0, there is no distinct cue qubit. Other partitions
are in the Supplement.

\subsubsection{Symmetry and robustness controls}
\label{sec:controls}

The task-generation procedure includes several symmetry controls that are
recorded per parameter setting. Each parameter setting is averaged over
\Nstream{} distinct stream pairs and
\Ncue{} distinct cue sequences in a crossed design, so that reported
performance and entanglement estimates are not tied to a single stream or
cue realization. To
diagnose any residual asymmetry between the two stream ports, each
parameter setting records
a port-imbalance quantity, the absolute difference in exact-readout decode
quality of the two streams \(u_1\) and \(u_2\); a symmetric architecture
should keep this near zero. The SI atlas regression uses that
imbalance as a mechanism control. Two stronger architecture-robustness checks
--- an explicit physical swap of the two stream ports and a reversal of the
injection order of the sub-step sequence [Eq.~\eqref{eq:statecycle}] --- are
reported in Supplementary Table~S1. All
main-text comparisons use the same predefined order.
The task is summarized in Fig.~\ref{fig:method}, and the resulting
performance map in Fig.~\ref{fig:atlas}.

\subsection{Entanglement computation}
\label{sec:ent-def}
In the main text \(\Ecut\) is the logarithmic negativity of one fixed
bipartition: the predeclared three-qubit input block \(\{0,1,2\}\) versus
the rest, \(E_N^{012|\overline{012}}\)
\cite{vidal2002computable,plenio2005logarithmic},
\begin{equation}
E_N=\log_2\lVert\hat\rho^{\Gamma_A}\rVert_1,
\label{eq:EN}
\end{equation}
with \(\Gamma_A\) the partial transpose on that block. A two-qubit
singlet has \(E_N=1\). For the \(012|\overline{012}\) cut, \(E_N\) is at
most \(\min(3,n-3)\) ebits (\(1\) at \(n=4\), \(2\) at \(n=5\),
and \(3\) at \(n\ge6\)). Main-text figures report this value in ebits,
not the per-ebit ratio \(E_N/\min(3,n-3)\). That ratio is the
three-qubit predictor in Supplementary Fig.~S7c, a mixed-stream size
scan, and is not an axis of Figs.~\ref{fig:sizesum} or~\ref{fig:deph}.
Qubit~\(0\) is the injection site; qubits \(1,2\) complete
the block. The quantity is a property of the post-injection state and
does not depend on \(N_M\). Other input-containing three-qubit cuts
are compared on the mixed-stream atlas in Supplementary Fig.~S5.
The one-qubit input cut is compared with this block in
Supplementary Fig.~S7c. Any apparent dependence of
entanglement on \(N_M\) comes only from the coupling that validation
selects at that \(N_M\).

\subsection{Coherence computation}
\label{sec:coh-def}
The second quantumness marker is the Baumgratz \(\ell_1\) coherence of
the full \(n\)-qubit state in the computational (\(\sigma^z\))
basis~\cite{baumgratz2014},
\begin{equation}
C_{\ell_1}(\hat\rho)=\sum_{i\neq j}\lvert\rho_{ij}\rvert
=\sum_{i,j}\lvert\rho_{ij}\rvert-\sum_{i}\lvert\rho_{ii}\rvert.
\label{eq:cl1}
\end{equation}
A state that is diagonal in this basis has \(C_{\ell_1}=0\). A uniform
superposition of all \(2^n\) computational-basis states has
\(C_{\ell_1}=2^n-1\) (\(15\) at \(n=4\), \(255\) at \(n=8\)). We report
that raw sum, not the Baumgratz ratio \(C_{\ell_1}/(2^n-1)\), because
the size survey covers \(n=4\)--\(8\). Populations do not enter; only
off-diagonal amplitudes do. The computational basis is the same product
basis used for readout, so \(C_{\ell_1}\) is a state diagnostic, not a
readout feature: the trained features \(\langle\hat\sigma^z_i\rangle\)
and \(\langle\hat\sigma^z_i\hat\sigma^z_j\rangle\) are diagonal. At
complete dephasing (\(\gamma_\varphi=1\)) the state is diagonal, so
\(C_{\ell_1}=0\).

Like \(E_N\), \(C_{\ell_1}\) is independent of \(N_M\) except through
which coupling \(J^\star(n,N_M)\) is held fixed. It is evaluated on the
correct-cue drive at four equally spaced times after washout, and
averaged over those snapshots.
The dephasing survey uses sixteen master seeds for \(C_{\ell_1}\)
(score and ranking \(E_N\) use \(\Nmaster\) masters). Fig.~\ref{fig:deph}
then reports the equal-cell-weight mean of that snapshot average over
the \(\Nsurvey\) survey cells. Because the two markers have different
ranges and use different numbers of master seeds, comparisons of their
curve shapes are qualitative.

All quantities reported in the main text are computed using a predefined
analysis pipeline, and all figures and reported numerical values are
generated from the resulting derived data tables. The \(n=8\)
performance map (Fig.~\ref{fig:atlas}) uses \(\Ncoup\) couplings,
\(\Nres\) reservoir realizations, \(\Nstream\) stream pairs,
\(\Ncue\) cue sequences, and seven values of \(N_M\) including exact
readout. Size scaling and the ranking survey
(Figs.~\ref{fig:sizesum} and~\ref{fig:deph}) use a separate
campaign: \(\Nmaster\) nested master seeds, \(\Nstreamscal\) stream
pairs, \(\Ncue\) cue sequences, \(n=\Nscalmin\)--\(\Nscalmax\), and
\(\gamma_\varphi\) on a \(\Ndamp\)-point dephasing ladder in \([0,1]\).

\subsection{Stream generation}
Each stream pair \((u_{1,t},u_{2,t})\) is a shared, reproducible draw generated
by the task-generation procedure. For the main parameter sweep the two
streams are independent i.i.d.\ uniform draws on \([-1,1]\) of length
\(T=1200\)
, each subsequently standardized to zero mean and unit variance so that neither
stream carries an amplitude signature of the cue. Main-text figures use
these unmixed streams (\(\alpha=0\)); mixed-stream (\(\alpha>0\)) controls
are defined by Supplementary Eq.~(S1). Stream pair \(s\) is
generated using seed \(821{,}000+s\); the same \(\Nstream\) pairs are reused
across every coupling, reservoir, and tested \(N_M\) value, so
performance differences are never confounded by input resampling. The same routine can drive the reservoir with two independent Mackey--Glass
series (delay \(\tau=30\)) instead of i.i.d.\ draws. That variant is a
supplementary sanity check (Supplementary Fig.~S10b), not the main
performance map. A conventional one-series forecast --- one Mackey--Glass
trajectory of delay \(\tau=17\), teacher-forced, with the target \(h\) steps
ahead --- is Supplementary Fig.~S10c.
One example of the target and the prediction for each Mackey--Glass
check is Supplementary Fig.~S11.

\subsection{Cue generation}
The cue \(c_t\in\{0,1\}\) (generated using seed \(822{,}000+q\)) is a block-structured, approximately balanced binary sequence
with persistence: \(c_0\) is a fair coin and, at each step, \(c_t=c_{t-1}\) with
probability \(0.90\)
\ and flips otherwise, producing persistent cue blocks rather than an i.i.d.\
label. The block structure creates sustained selection contexts and avoids
unrealistically rapid cue switching. The cue remains available at
every step, so the task does not require cue memory. It is still
nontrivial for the tuned linear baseline because the target contains
the bilinear cue--stream product. We use \(\Ncue\) distinct cue sequences.

\subsection{Sequential substep input injection}
The resets in Eq.~\eqref{eq:statecycle} are sequential substep input injection.
Readout-time multiplexing is a separate operation: it rereads one logical step
and does not change how the inputs enter. All three signals enter through
qubit~\(0\), so \(\Ecut\) is an input-block cut. The order of those resets
is an architectural choice.
The port-swap and order-reversal controls are Supplementary Table~S1.

\subsection{Reservoir Hamiltonian and time step}
Each reservoir is a fixed \(\Nqubits\)-spin system with the random
transverse-field XX Ising Hamiltonian of Eq.~\eqref{eq:H},
with transverse field \(h=1\) and symmetric couplings \(J_{ij}\) drawn once per
reservoir seed from \(\mathrm{Unif}(-J_s,J_s)\); the control parameter is the
coupling ratio \(J_s/h\). The performance map uses the \(\Ncoup\)-point coupling grid
\(J_s/h=10^{k/10}\), \(k=-7,\dots,5\) (spanning \(\approx0.20\) to \(\approx3.16\)
in \(0.1\)-decade increments).
Six of these coupling values match an earlier simulation and are kept
as a consistency
check. Two hundred reservoir realizations (\(r=0,\dots,199\)) use one
seed sequence. The first fifty match an earlier Hamiltonian set; the
rest continue that sequence. The fixed-\(n=\Nqubits\)
parameter sweep is closed-system: dephasing is
disabled (\(\gamma_\varphi=0\)); it enters only in the separate
size-and-dephasing numerical study.

\subsection{Finite number of measurements}
\label{sec:finite-nm}
Exact readout uses the reservoir's diagonal populations
\(P_t=\mathrm{diag}(\hat\rho_t)\). A finite number of measurements replaces \(P_t\) with empirical
bitstring frequencies drawn by multinomial sampling. The finite values are
\(N_M\in\{100,300,10^3,3\times10^3,10^4,3\times10^4\}\).
\(N_M=\infty\) denotes infinitely many measurements, that is, exact expectation values.
The finite values are obtained by \emph{nested} sampling: for each draw the
multinomial counts are accumulated in increments across the grid
(\(c_{100}\sim\mathrm{Mult}(100,p_t)\),
\(c_{300}=c_{100}+\mathrm{Mult}(200,p_t)\), and so on). The samples
are nested such that each larger sample contains all observations from
the corresponding smaller sample, so this construction measures the
marginal value of additional measurements rather than independent experiments. Eight
independent nested draw sequences per cue condition are generated, each sequence is scored
separately, and the eight \emph{scores} are averaged;
features from different draws are never combined, so every reported score is the
Monte-Carlo mean of the score \emph{at} \(N_M\), not a score at an
inflated effective number of measurements. A supplementary dense-grid parameter sweep uses
the quarter-decade grid \(N_M\in\{10^{3.5},10^{3.75},\ldots,10^{5}\}\). Five
reservoir simulations did not complete because of a computational failure
(identifiers \(12,52,92,132,172\)); these simulations were excluded uniformly
from the dense-grid analysis, independently of their outcomes, so that sweep
is analyzed as a uniform \(195\)-reservoir ensemble and is used only for the
onset refinement reported in the text.
Figure~\ref{fig:atlas}c continues the same
nested multinomial sequence to \(N_M=10^{5}\) for visualization only, reusing the same state
trajectories and eight nested draws; that larger \(N_M\) value does not enter
selection of the operational success criterion, coupling selection, or the
entanglement--performance analysis.

Two idealizations of this measurement model should be stated explicitly. First,
sampling is applied to \(\mathrm{diag}(\hat\rho_t)\) with no measurement back-action
on the reservoir state: physically this corresponds to a restart protocol in
which the input prefix is re-prepared to read the reservoir state at each step.
Two costs then follow. The number of \emph{measured} circuit instances is
\(N_M\,T=1.2\times10^{7}\) at \(N_M=10^{4}\), \(T=1200\). Because each state must
be reached by replaying its input prefix from the start, the \emph{cumulative}
number of logical-step executions is
\(N_M\sum_{t=1}^{T}t=N_M\,T(T{+}1)/2=7.206\times10^{9}\), the
replay depth incurred by measuring the state at every step without measurement
back-action. The quoted \(N_M\) is therefore the number of measurements per step,
rather than the total experimental repetition count. This restart-based cost estimate is an \emph{optimistic
simulation-level lower bound}, because errors in preparing and measuring
the state, in the reset, in the gates, and from drift are left out and would raise
the effective cost. The second idealization is that all \(d(n)\) observables are estimated from the same
bitstrings (one measurement setting), which is what makes the sampling errors
correlated across features. Both idealizations are shared by all parameter
settings, so they
shift the absolute scale of the number of measurements rather than any comparison made in this paper.
The numerical thresholds in the Results should therefore be read as an
optimistic simulation-level lower bound. The Discussion returns to
what these thresholds mean as a hardware cost.

\subsection{Feature map}
The readout features are the one- and two-body \(\sigma^z\) observables of the sampled
distribution: the \(\Nqubits\)
single-qubit means \(\langle \hat\sigma^z_i\rangle\) and all \(\binom{\Nqubits}{2}\)
pairwise \(\langle \hat\sigma^z_i\hat\sigma^z_j\rangle\) moments, evaluated from the empirical counts at
each \(N_M\). All features are diagonal, so the exact and finite-\(N_M\) feature maps
share an identical definition and differ only in estimator noise.

\subsection{Ridge readout, scoring, and numerical studies}
\label{sec:methods-metrics}
\label{sec:sizedamp}
A single linear readout is trained jointly across both cue values by ridge
regression. The time series is partitioned after a washout of the first \(100\)
\ steps: \(70\%\) of the post-washout steps form the training window and the
remaining \(30\%\) the test window; within the training window the last \(20\%\)
is held out for validation. Two copies of one reservoir that differ only
in the initial state, and that see the same inputs, have trace distance
at most \(3.5\times 10^{-10}\) by the end of this washout
(Supplementary Fig.~S10a). The ridge penalty is selected on the validation
window from
\(\lambda\in\{10^{-5},10^{-4},10^{-3},10^{-2},10^{-1},1\}\)
\ and never tuned on the test window. Throughout, the reported score is
\(S=100R^2\) on the test window, where \(R\) is the Pearson correlation
between the readout prediction \(\tilde y_t\) and the
target \(y_t\). Perfect prediction scores \(100\); predicting the
test-set mean scores \(0\). A valid-cue drive and a shuffled-cue control (block-preserving
circular shift of the cue) are scored separately. A parameter setting meets an \emph{operational success criterion} when the reservoir outperforms the linear baseline
(\(\directbase\) on the atlas ensemble; \(\directscal\) on the independent
size-survey dataset) and the valid-cue score exceeds the shuffled-cue score.
The criteria defining the
operating regime and the atlas regression analysis of entanglement and
performance are in the supplementary Additional Methods.

Figs.~\ref{fig:sizesum} and~\ref{fig:deph} come from a paired
size-scaling study:
\(\Nmaster\) master seeds, nested \(n=4,\ldots,8\) principal
submatrices sharing the coupling structure and input qubit,
\(N_M\in\{10^{3},10^{4},10^{5},10^{6},\infty\}\) on the ranking grid,
and dephasing \(\gamma_\varphi\) on a \(\Ndamp\)-point ladder in
\([0,1]\).
For both figures, \(J^\star(n,N_M)\) is the coupling that
maximizes mean score at \(\gamma_\varphi=0\); the same cells are scored
at every dephasing rung. Fig.~\ref{fig:sizesum} reports ensemble means
versus \(n\) at \(\gamma_\varphi=0\) and the smallest tested \(N_M\) at
which that mean exceeds the linear baseline. Fig.~\ref{fig:deph}
reports equal-cell-weight means of score, \(E_N\), and \(C_{\ell_1}\).
Error bars on the equal-cell-weight means are the standard error
across the \(\Nsurvey\) cells. They reflect differences among
conditions, not repeated runs of one setting. The dephasing comparison at
\(\gamma_\varphi=\gdephMax\) is the paired difference of the same cell
means relative to \(\gamma_\varphi=0\).

\section*{Acknowledgments}
We thank Terrence C. Stewart for useful discussions. This work was supported by the National Research Council through its Applied Quantum Computing Challenge Program and the Natural Sciences and Engineering Research Council (NSERC) of Canada through its NSERC Discovery Grant Program.

\section*{Conflict of interest}
H.Z.-H. is a founder of QNeura - Quantum Artificial Intelligence Advanced Technology Inc. The company did not fund or play any other role in this study. C.S. declares no conflict of interest.

\section*{Code availability}
The source code supporting the findings of this study is available at \url{https://github.com/hadizh20/qrcatt}.

\bibliography{refs}
\clearpage
\onecolumngrid
\pdfximage{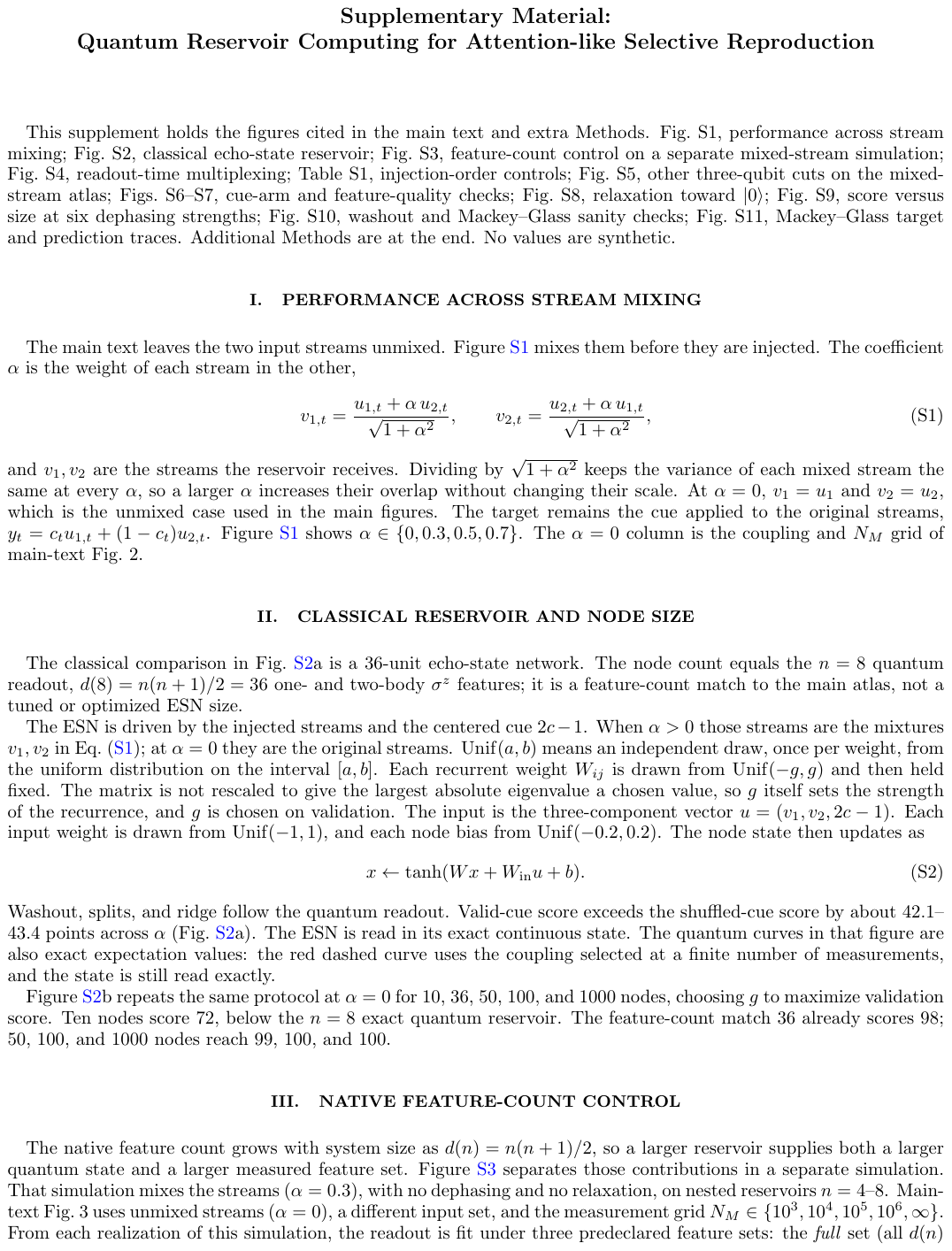}
\newcount\sipage
\sipage=1
\loop


\section*{Supplementary material}

# Supplementary Material:
# Quantum Reservoir Computing for Attention-like Selective Reproduction

This supplement holds the figures cited in the main text and extra Methods. Fig. S1, performance across stream mixing; Fig. S2, classical echo-state reservoir; Fig. S3, feature-count control on a separate mixed-stream simulation; Fig. S4, readout-time multiplexing; Table S1, injection-order controls; Fig. S5, other three-qubit cuts on the mixed-stream atlas; Figs. S6–S7, cue-arm and feature-quality checks; Fig. S8, relaxation toward $|0\rangle$; Fig. S9, score versus size at six dephasing strengths; Fig. S10, washout and Mackey–Glass sanity checks; Fig. S11, Mackey–Glass target and prediction traces. Additional Methods are at the end. No values are synthetic.

## I. PERFORMANCE ACROSS STREAM MIXING

The main text leaves the two input streams unmixed. Figure S1 mixes them before they are injected. The coefficient $\alpha$ is the weight of each stream in the other,

$$v_{1,t} = \frac{u_{1,t} + \alpha\, u_{2,t}}{\sqrt{1+\alpha^2}}, \qquad v_{2,t} = \frac{u_{2,t} + \alpha\, u_{1,t}}{\sqrt{1+\alpha^2}}, \tag{S1}$$

and $v_1, v_2$ are the streams the reservoir receives. Dividing by $\sqrt{1+\alpha^2}$ keeps the variance of each mixed stream the same at every $\alpha$, so a larger $\alpha$ increases their overlap without changing their scale. At $\alpha = 0$, $v_1 = u_1$ and $v_2 = u_2$, which is the unmixed case used in the main figures. The target remains the cue applied to the original streams, $y_t = c_t u_{1,t} + (1 - c_t)u_{2,t}$. Figure S1 shows $\alpha \in \{0, 0.3, 0.5, 0.7\}$. The $\alpha = 0$ column is the coupling and $N_M$ grid of main-text Fig. 2.

## II. CLASSICAL RESERVOIR AND NODE SIZE

The classical comparison in Fig. S2a is a 36-unit echo-state network. The node count equals the $n = 8$ quantum readout, $d(8) = n(n+1)/2 = 36$ one- and two-body $\sigma^z$ features; it is a feature-count match to the main atlas, not a tuned or optimized ESN size.

The ESN is driven by the injected streams and the centered cue $2c-1$. When $\alpha > 0$ those streams are the mixtures $v_1, v_2$ in Eq. (S1); at $\alpha = 0$ they are the original streams. $\mathrm{Unif}(a, b)$ means an independent draw, once per weight, from the uniform distribution on the interval $[a, b]$. Each recurrent weight $W_{ij}$ is drawn from $\mathrm{Unif}(-g, g)$ and then held fixed. The matrix is not rescaled to give the largest absolute eigenvalue a chosen value, so $g$ itself sets the strength of the recurrence, and $g$ is chosen on validation. The input is the three-component vector $u = (v_1, v_2, 2c - 1)$. Each input weight is drawn from $\mathrm{Unif}(-1, 1)$, and each node bias from $\mathrm{Unif}(-0.2, 0.2)$. The node state then updates as

$$x \leftarrow \tanh(Wx + W_{\mathrm{in}}u + b). \tag{S2}$$

Washout, splits, and ridge follow the quantum readout. Valid-cue score exceeds the shuffled-cue score by about 42.1–43.4 points across $\alpha$ (Fig. S2a). The ESN is read in its exact continuous state. The quantum curves in that figure are also exact expectation values: the red dashed curve uses the coupling selected at a finite number of measurements, and the state is still read exactly.

Figure S2b repeats the same protocol at $\alpha = 0$ for 10, 36, 50, 100, and 1000 nodes, choosing $g$ to maximize validation score. Ten nodes score 72, below the $n = 8$ exact quantum reservoir. The feature-count match 36 already scores 98; 50, 100, and 1000 nodes reach 99, 100, and 100.

## III. NATIVE FEATURE-COUNT CONTROL

The native feature count grows with system size as $d(n) = n(n+1)/2$, so a larger reservoir supplies both a larger quantum state and a larger measured feature set. Figure S3 separates those contributions in a separate simulation. That simulation mixes the streams ($\alpha = 0.3$), with no dephasing and no relaxation, on nested reservoirs $n = 4$–8. Main-text Fig. 3 uses unmixed streams ($\alpha = 0$), a different input set, and the measurement grid $N_M \in \{10^3, 10^4, 10^5, 10^6, \infty\}$. From each realization of this simulation, the readout is fit under three predeclared feature sets: the *full* set (all $d(n)$

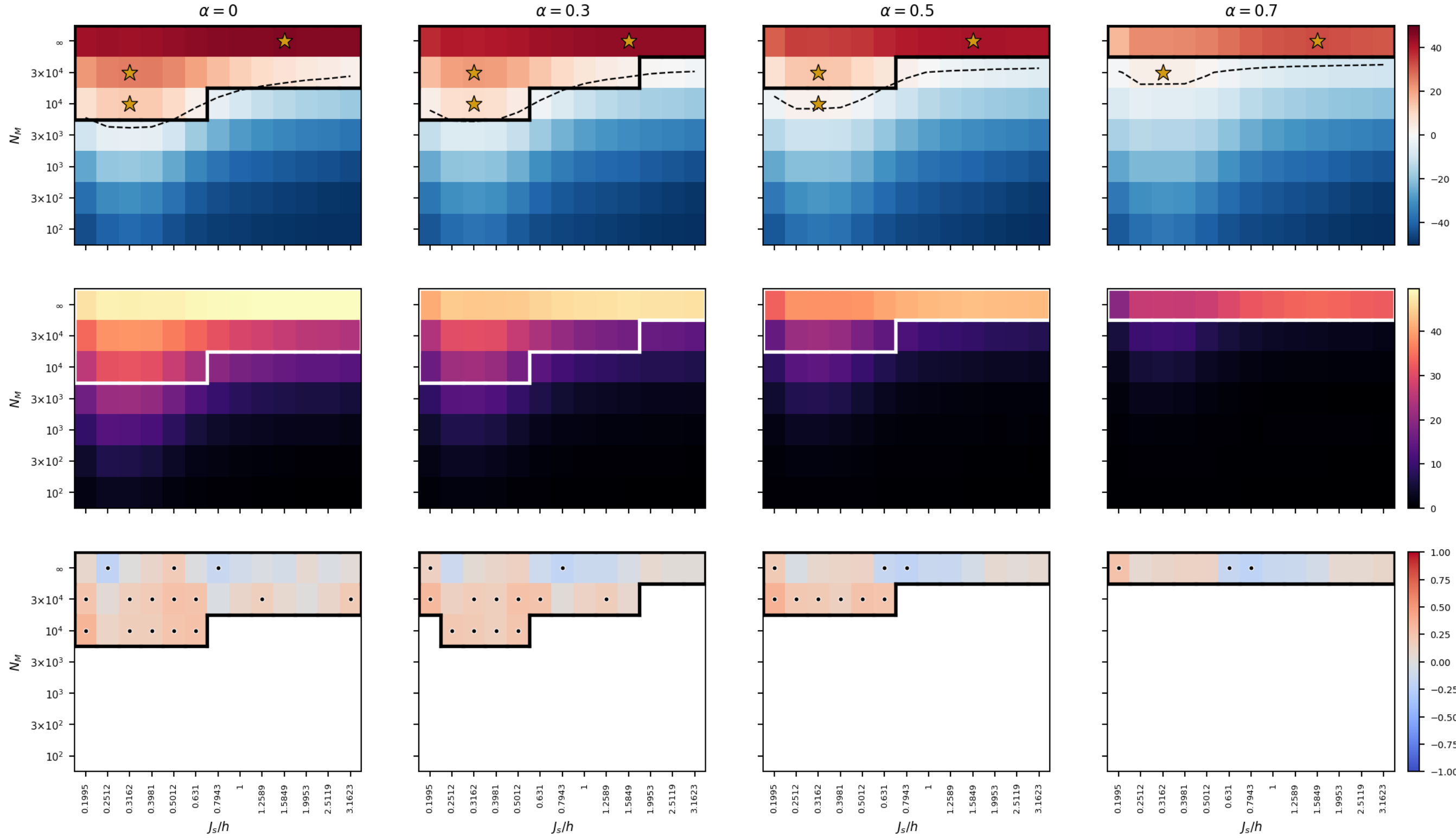


FIG. S1. **Performance across stream mixing, at a finite number of measurements.** Columns are stream mixing $\alpha \in \{0, 0.3, 0.5, 0.7\}$. The $\alpha = 0$ column uses the same coupling and $N_M$ grid as main-text Fig. 2, where the linear baseline is 50.7. Fig. 2 is the valid-cue score on that grid. The rows, from top to bottom, are the score minus the linear baseline, the valid-cue score minus the shuffled-cue score, and the correlation of entanglement with that gap. Within each panel the horizontal axis is the internal coupling $J_s/h$ and the vertical axis is $\log_{10} N_M$ (exact readout as the separate top category). Stars mark the validation-selected coupling $J^*(\alpha, N_M)$. The thick outline marks settings that pass the score checks in the Additional Methods. The dashed curve is where the valid-cue score equals the linear baseline. In the bottom row, cells outside that outline are blank, cells whose entanglement barely varies are grey, and a dot marks cells whose 95% interval excludes zero.

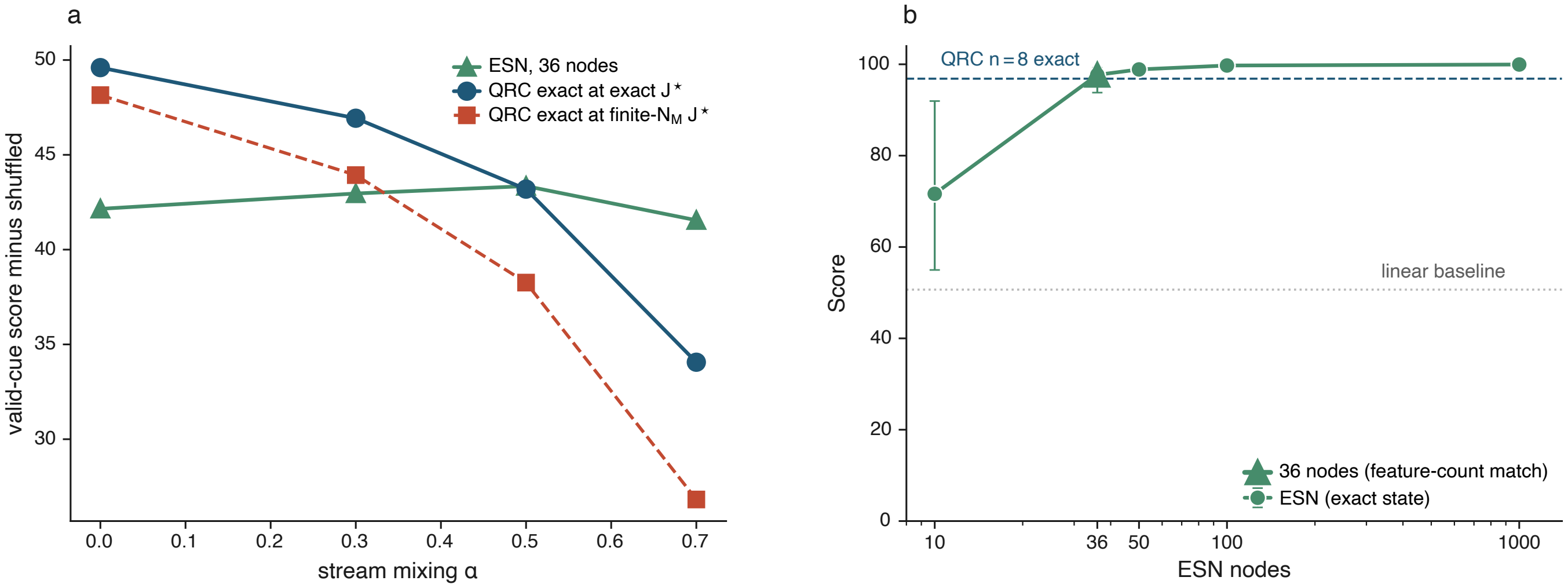


FIG. S2. **Classical echo-state reservoir.** **(a)** Valid-cue score minus the shuffled-cue score, versus stream mixing, exact-state readout. Green: 36-unit ESN, matching the $n = 8$ quantum feature count $d(8) = 36$. Navy: quantum reservoir at its exact-optimal coupling. Red dashed: same quantum reservoir at the coupling selected at a finite number of measurements, still read exactly. **(b)** Valid-cue score versus ESN size at $\alpha = 0$. Each point uses the $g$ that maximizes validation score. Bars are 2.5–97.5 percentiles over ESN seeds. Triangle: the feature-count match 36. Dotted: linear baseline. Dashed: $n = 8$ QRC with exact expectation values. The ESN state is read exactly; this is not a resource-matched quantum–classical test.

one- and two-body $\sigma^z$ observables), a *one-body-only* set (the $n$ observables $\langle\hat{\sigma}_i^z\rangle$), and a *fixed-count* set (the first $10 = d(4)$ observables in the declared order $\langle\hat{\sigma}_0^z\rangle, \ldots, \langle\hat{\sigma}_{n-1}^z\rangle, \langle\hat{\sigma}_0^z\hat{\sigma}_1^z\rangle, \ldots$, so the feature *count* is held fixed while the state grows). Only the readout column subset differs within the simulation, so the three readouts share that realization's quantum state.

Figure S3 shows the outcome. The two-body correlators carry the selection: the one-body-only readout scores far below the full readout at every size, confirming that the cue–stream interaction is encoded in the $\langle\hat{\sigma}_i^z\hat{\sigma}_j^z\rangle$ features. The fixed-count readout lies between the two, so in this simulation a substantial part of the exact-readout gain with size reflects the growing measured feature set rather than the growing state alone. Isolating the finite-$N_M$ covariance burden would also require matching the resolvable feature rank, which this count-matched control leaves free.

## IV. READOUT-TIME MULTIPLEXING

Input injection is unchanged. After the boundary state of a logical step we free-evolve a copy and read $V$ equally spaced times (virtual nodes) [1], $\hat{\rho}_k = \hat{U}_{\rm sub}^k \hat{\rho}\,(\hat{U}_{\rm sub}^k)^\dagger$ with $\hat{U}_{\rm sub} = e^{-i\hat{H}\tau/V}$. Node 0 is the main-text readout ($V = 1$); the feature vector then has length $V\,d(n)$. We test $V \in \{1, 5, 10\}$ at $n \in \{6, 8\}$, $N_M = 10^4$, with no dephasing or relaxation.

Figure S4 is the control cited in the main text. With the total number of measurements held fixed, each node gets $N_M/V$ measurements: at $n = 6$ the score rises from 74 to 81 at $V = 5$ and then saturates; at $n = 8$ it is flat (62). Giving every node the full $N_M$ raises the $n = 8$ score from 62 to 85 at $V = 10$, but spends $V$ times more measurements. Virtual nodes are therefore an extra measurement resource, not a way around the finite-sampling limitation.

## V. INJECTION-ORDER AND PORT-SWAP CONTROLS

The two input streams and the cue are written, one after another, onto the same input qubit. The predefined order writes stream 1, then stream 2, then the cue, so stream 1 is partly overwritten by the two later writes and the cue is the last value written before the measurement. A port swap keeps that rhythm and exchanges the two streams, writing stream 2 first. An order reversal writes the cue first and the streams last.

The test was run at $\alpha = 0.3$. Each injected value is then a mixture of both streams, Eq. (S1), so the two writes carry different mixtures. Exchanging the streams changes which mixture is written first. The main text uses $\alpha = 0$. There each stream is injected by itself: the first write is $u_1$ and the second is $u_2$, with none of the other stream mixed in. Those are two different noise sequences with the same mean and variance. A swap would exchange those two sequences and change which one is written first. The main-text figures keep the predefined order, and this table does not re-estimate those scores.

We re-ran the $n = 8$, $\alpha = 0.3$ atlas slice (13-point coupling grid, 24 reservoirs, 4 stream pairs $\times$ 4 cue sequences) under the port swap $(v_2, v_1, c)$ and the order reversal $(c, v_2, v_1)$, with the total step time fixed. Table S1 reports the validation-selected operating points.

Swapping the stream ports leaves the score unchanged. The paired change in valid-cue score against the base order is $+0.2 \pm 2.5$ (mean $\pm$ SD over 4992 paired drives at exact readout), and the block negativity changes by $-0.0005 \pm 0.009$. Reversing the injection order changes the exact-readout score by $-1.1 \pm 3.1$. At $N_M = 10^4$ the reversed schedule exceeds the linear baseline by 24.7 points, against 10.2 for the predefined order. Moving the stream injections closer to the measurement is a possible explanation for that improvement, and the mechanism remains open.

## VI. CUT-ARBITRARINESS SWEEP

Figure S5 repeats the fitted entanglement slope on every other three-qubit cut of the mixed-stream atlas. The number is the change in score, in points, for a one-standard-deviation rise in the entanglement of that cut. It is not the dephasing curves of main-text Fig. 4 repeated for other partitions. The predeclared three-qubit input-containing block $\{0, 1, 2\}$ is one of 21 three-qubit subsets containing the input qubit. Repeating the valid-cue drive at every operating-region $(J, \alpha)$ pair (32 post-washout snapshots, the original 50 reservoirs) and estimating the raw cross-fitted marker for each of the 21 cuts under the identical model yields the sweep in Fig. S5: every cut is resolved, in a narrow band from $+0.79$ to $+1.25$, and the predeclared block ranks 3rd of 21. The association is a diffuse property of input-containing blocks rather than a feature of the specific partner qubits; feature-quality-adjusted estimates for the extreme cuts ($+0.53$ and $+0.32$) bracket the predeclared block's adjusted value.

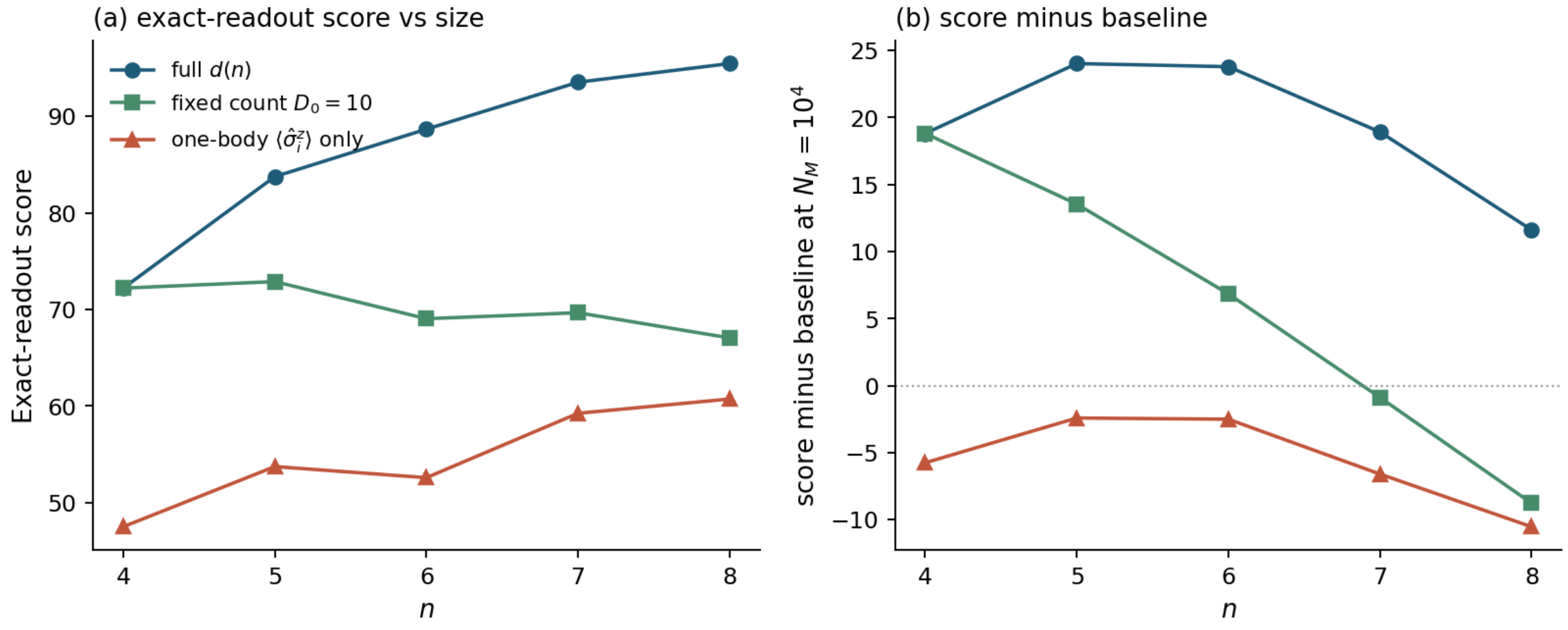


FIG. S3. **Native feature-count control** (separate simulation; $\alpha = 0.3$, no dephasing or relaxation). Main-text Fig. 3 is the unmixed survey ($\alpha = 0$) on a different input set. (a) Exact-readout valid-cue score versus system size $n$ for the full $d(n)$, one-body-only, and fixed-count (10) readouts. (b) Score minus the linear baseline, at $N_M = 10^4$, for the same three readouts. The one-body-only readout scores well below the full readout at every size; the fixed-count readout lies between them.

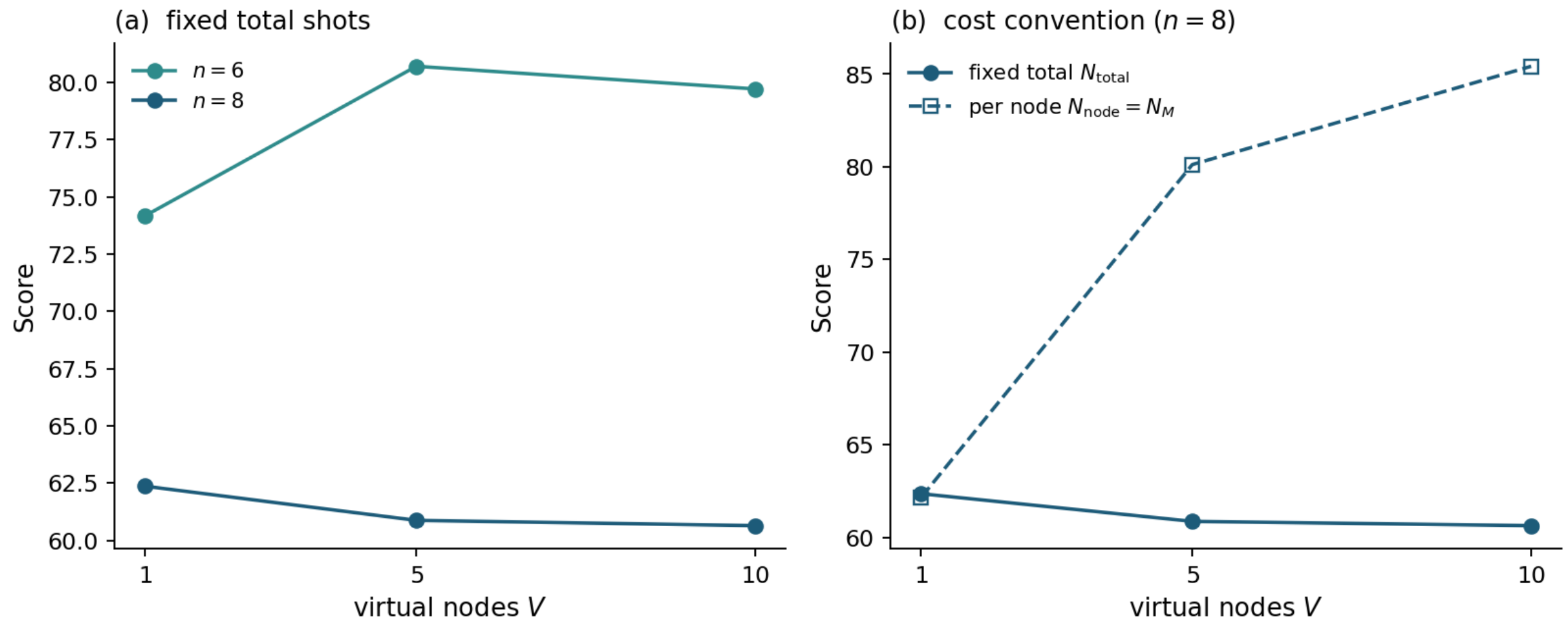


FIG. S4. **Readout-time multiplexing** ($N_M = 10^4$, no dephasing or relaxation). **(a)** Score versus virtual nodes $V$ at fixed total number of measurements, for $n = 6, 8$. **(b)** Fixed-total versus per-node cost at $n = 8$; the per-node curve spends $V$ times more measurements.

TABLE S1. **Injection-order and port-swap controls** ($n = 8$, $\alpha = 0.3$, 24 reservoirs). Validation-selected operating points per schedule at exact readout and $N_M = 10^4$. "Above baseline" is the valid-cue score minus the linear baseline 51.4 on these $\alpha = 0.3$ inputs. "Valid minus shuffled" is the valid-cue score minus the shuffled-cue score.

| schedule | $N_M$ | $J^\star$ | valid-cue score | above baseline | valid minus shuffled | $E_N$ |
|---|---|---|---|---|---|---|
| base $(v_1, v_2, c)$ | $\infty$ | 2.5119 | 95.4 | +44.0 | 45.9 | 0.482 |
| swap $(v_2, v_1, c)$ | $\infty$ | 1.2589 | 95.9 | +44.5 | 46.4 | 0.522 |
| reversed $(c, v_2, v_1)$ | $\infty$ | 1.5849 | 94.6 | +43.2 | 43.9 | 0.505 |
| base $(v_1, v_2, c)$ | $10^4$ | 0.2512 | 61.6 | +10.2 | 23.1 | 0.338 |
| swap $(v_2, v_1, c)$ | $10^4$ | 0.2512 | 63.5 | +12.1 | 24.3 | 0.337 |
| reversed $(c, v_2, v_1)$ | $10^4$ | 0.3162 | 76.1 | +24.7 | 30.0 | 0.499 |

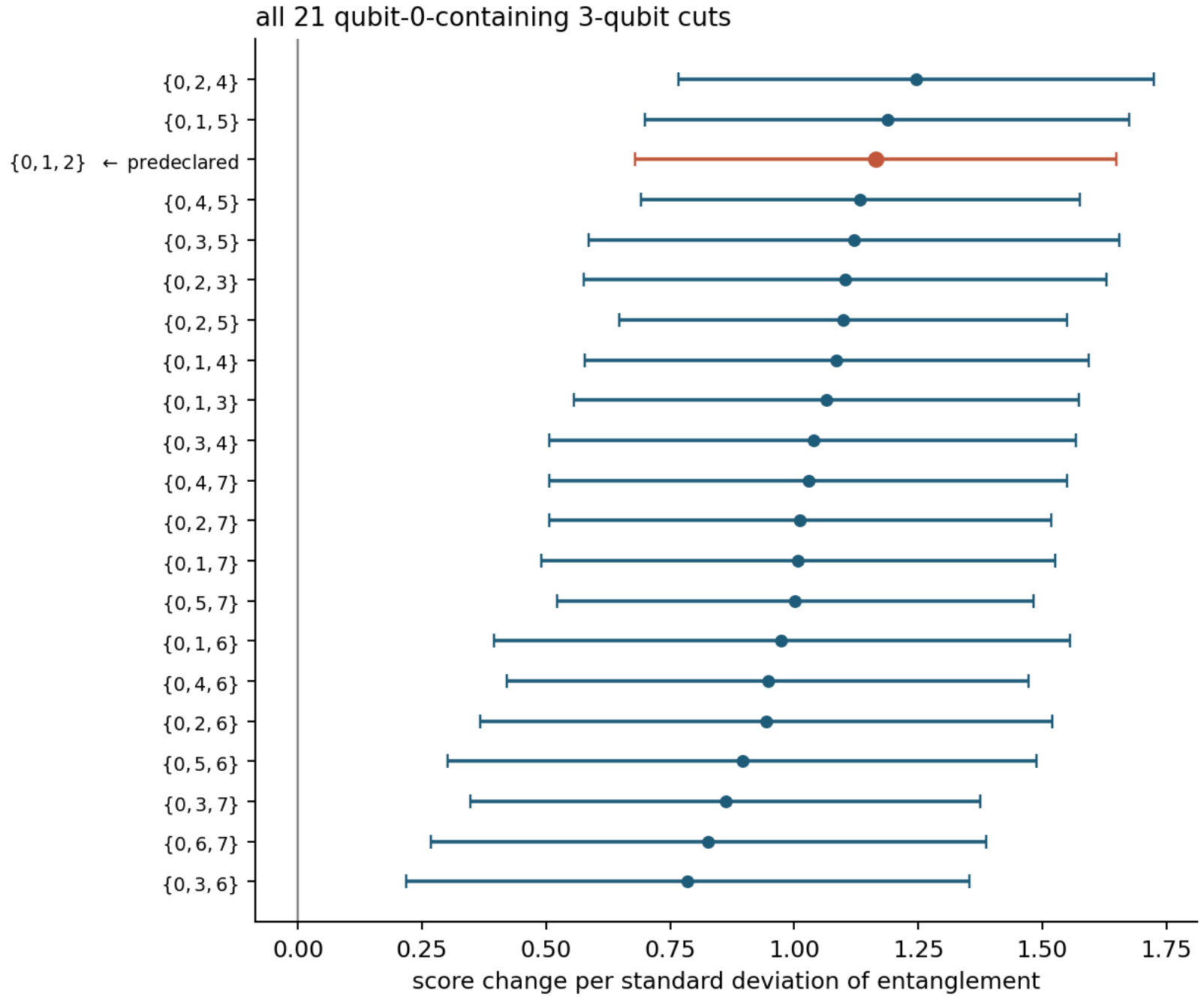


FIG. S5. **Cut-arbitrariness sweep on the mixed-stream atlas.** Change in score per standard deviation of entanglement, for all 21 three-qubit cuts containing the input qubit (sorted; 95% intervals), from the 32-snapshot simulation of the operating region on the original 50 reservoirs. This is the atlas association, not the dephasing curves of main-text Fig. 4. The predeclared $\{0, 1, 2\}$ block (red) ranks 3rd of 21; all cuts are resolved within a narrow band.

## VII. ENTANGLEMENT AND SCORE WITH THE VALID AND THE SHUFFLED CUE

Figs. S6 and S7 compare reservoirs inside a successful setting and report how their score changes with entanglement. A setting is one coupling $J_s/h$, one stream mixing $\alpha$, and one number of measurements $N_M$. It is successful when the scores pass the checks in the Additional Methods. The same reservoirs are not used both to pick those settings and to measure the slope. The 200 reservoirs are split in half, $r < 100$ and $r \geq 100$. One half decides which settings are successful. The slope is fit only on the other half. The two halves then swap roles, and the two slopes are averaged.

Reservoirs with more entanglement have a higher score on both ways of applying the cue. A cue arm is one of those ways. The valid arm uses the cue that selects the target stream. The shuffled arm uses that same cue sequence with its timing scrambled: the cue is still injected and still drives the reservoir, but it no longer says which stream to copy. Reservoirs that share a setting are compared with one another: the mean score and the mean entanglement of the setting are removed, and the overall means are put back, so the axes stay in score points and in $E_N$. The marker is the predeclared input-block negativity $E_N^{012|\overline{012}}$. The main text does not plot these slopes.

Figure S6a draws both arms. Binned means of the valid-cue score and of the shuffled-cue score both rise with entanglement.

Figure S6b is that rise as a slope, the change in score for a one-standard-deviation rise in entanglement. Open marks are the raw fit. Filled marks also adjust for feature quality. The shuffled-cue slope $+0.90\,[+0.68, +1.12]$ points per standard deviation is about 66% of the valid-cue slope $+1.37\,[+1.05, +1.69]$ (raw 46%). Their difference, $+0.47\,[+0.29, +0.65]$, agrees with the slope fitted directly to the valid-cue score minus the shuffled-cue score (to $< 5 \times 10^{-3}$). Most of the higher score is therefore present even when the cue is scrambled. The smaller difference is the part that depends on the cue being valid.

*a. Which cue arm supplies the entanglement.* The slopes above use the negativity recorded on the valid-cue drive, including for the shuffled-cue arm. A check drives 40 reservoirs at two couplings with the shuffled cue and recomputes the input-block negativity. The two drives give almost the same values: $r = 0.996$ after the same within-setting comparison (raw per-reservoir $r = 0.998$; mean bias $+1.6\%$, valid-cue negativity marginally higher). For these reservoirs, the slopes do not depend on which cue arm supplies the negativity. The check covers only these 40 reservoirs, so it is exploratory.

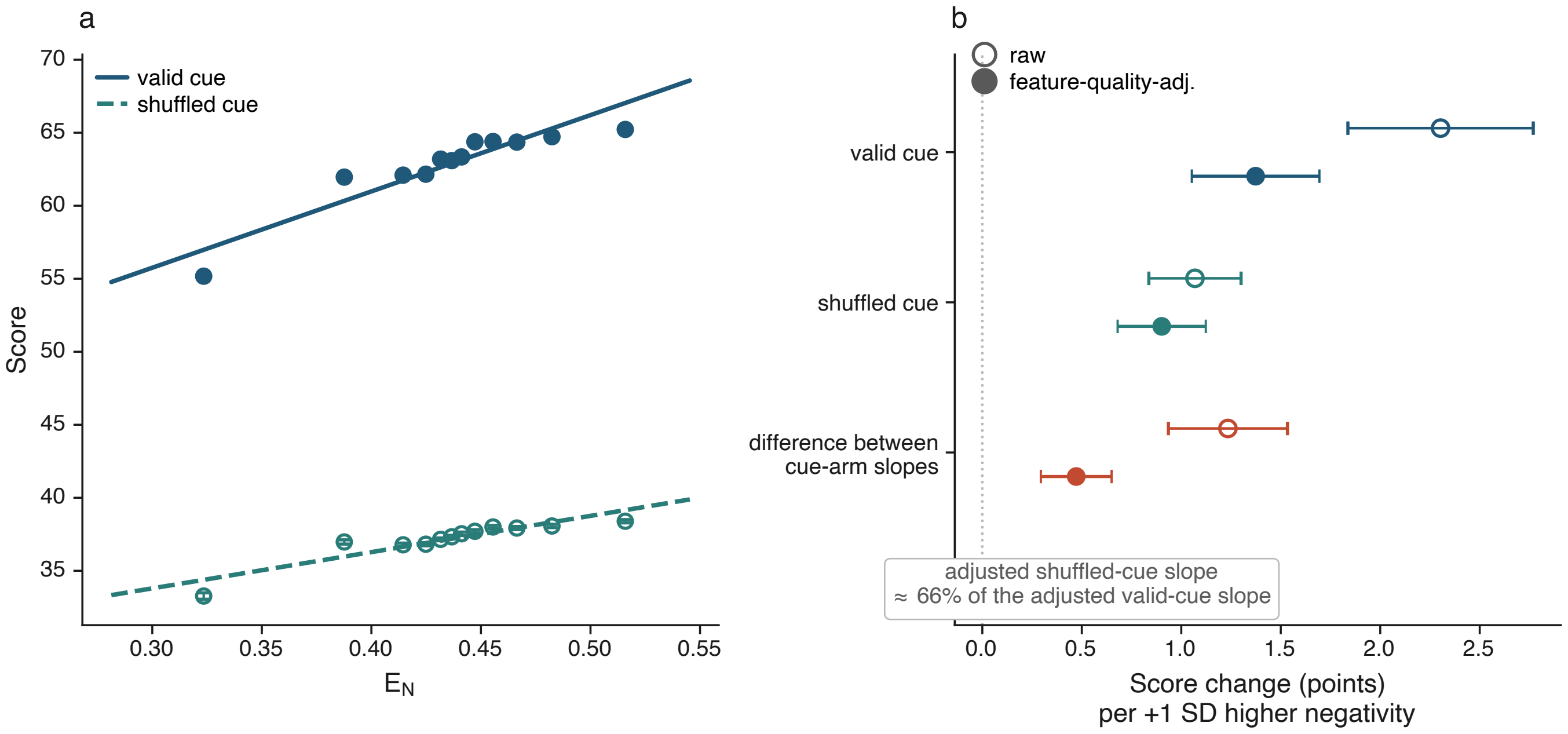


FIG. S6. **Entanglement and score with the valid and the shuffled cue.** Entanglement is the predeclared input-block negativity $E_N^{012|\overline{012}}$ (not the single-qubit cut). Both panels use only successful settings at a finite number of measurements, with the held-out split described in the text. Most of the rise is shared by the two cue arms; the part that depends on the cue being valid is the smaller difference. **(a)** Valid-cue and shuffled-cue scores versus $E_N$, comparing reservoirs that share a setting (binned means $\pm$SE). **(b)** Change in score per standard deviation of entanglement: valid cue, shuffled cue, and their difference; raw (open) and feature-quality-adjusted (filled).

## VIII. SCORE GAP BETWEEN THE CUE ARMS: CONTROLS AND CHOICE OF QUBITS

Figure S7 uses the same comparison and the same held-out split as Fig. S6. Here the score is the gap between the two cue arms, the valid-cue score minus the shuffled-cue score. Reservoirs that share a setting (the coupling $J_s/h$, the stream mixing $\alpha$, and the number of measurements $N_M$) are compared with each other after the mean of the setting is removed. These panels are separate from the dephasing curves of main-text Fig. 4.

Panel (a) shows a larger gap for reservoirs with more entanglement (binned means; slope +28 points per unit of $E_N^{012|\overline{012}}$). Figure S6a shows the two scores that make up this gap, each drawn on its own.

Panel (b) follows that rise through other differences among the reservoirs. Feature quality absorbs roughly 62% of the raw slope. A small positive remainder stays, and adding the coupling norm and the level spacing leaves that remainder in place.

Panel (c) repeats that remainder at each system size and for two choices of which qubits define the entanglement. This panel is a mixed-stream scan ($\alpha = 0.3$), separate from the unmixed survey in main-text Fig. 4. Navy is the three-qubit input block $\{0, 1, 2\}$, with predictor $E_N / \min(3, n-3)$. Red is the one-qubit input cut $\{0\}$. The adjusted slope is unresolved at $n \leq 5$, grows for $n \geq 6$, and is slightly larger for the block.

## IX. RELAXATION TOWARD $|0\rangle$

Main-text Figs. 3 and 4 use local dephasing. Figure S8 places the same closed-system $J^\star(n, N_M)$ cells on a dissipation ladder toward $|0\rangle$ (dephasing off). Moderate relaxation raises mean score while mean $E_N$ falls; only complete damping destroys selection.

## X. SCORE VERSUS SIZE AT SIX DEPHASING STRENGTHS

Main-text Fig. 3a is score versus size at $\gamma_\varphi = 0$; Fig. 3b is the onset heatmap. Figure S9 repeats the score-versus-size curve once for each of six dephasing strengths, $\gamma_\varphi = 0, 0.01, 0.1, 0.3, 0.6, 1$. Panel (a) is main-text Fig. 3a. The other

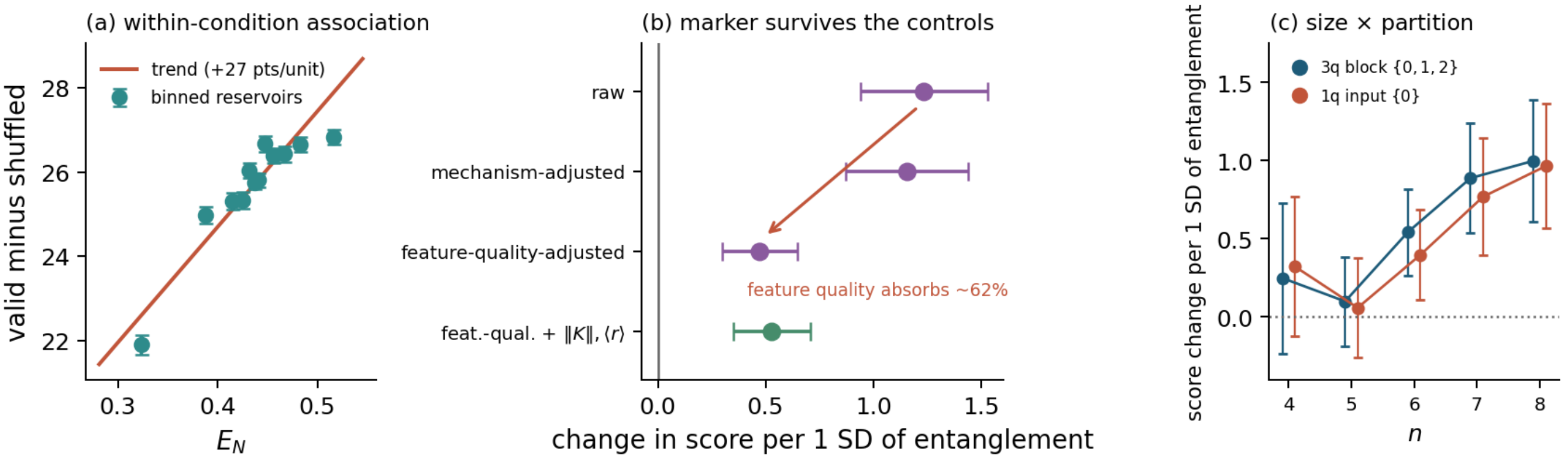


FIG. S7. **Score gap between the cue arms: controls and choice of qubits** (settings that pass the score checks in the Additional Methods). (a) Valid-cue score minus the shuffled-cue score versus $E_N^{012|\overline{012}}$, among reservoirs that share a coupling, a stream mixing, and a number of measurements (binned reservoirs; fitted slope +28 points per unit of entanglement). Fig. S6a draws the valid-cue score and the shuffled-cue score separately; this panel draws only their difference. (b) The same fitted slope across successive control sets—raw, mechanism-adjusted, and feature-quality-adjusted—plus the coupling-norm and level-spacing controls (green); the arrow marks the $\sim 62\%$ absorbed by feature quality. (c) That adjusted slope versus system size $n$ on the mixed-stream size scan ($\alpha = 0.3$), separate from main-text Fig. 4. Navy: three-qubit block $\{0, 1, 2\}$, with predictor $E_N / \min(3, n-3)$. Red: one-qubit input cut $\{0\}$.

panels are the same cells with dephasing on.

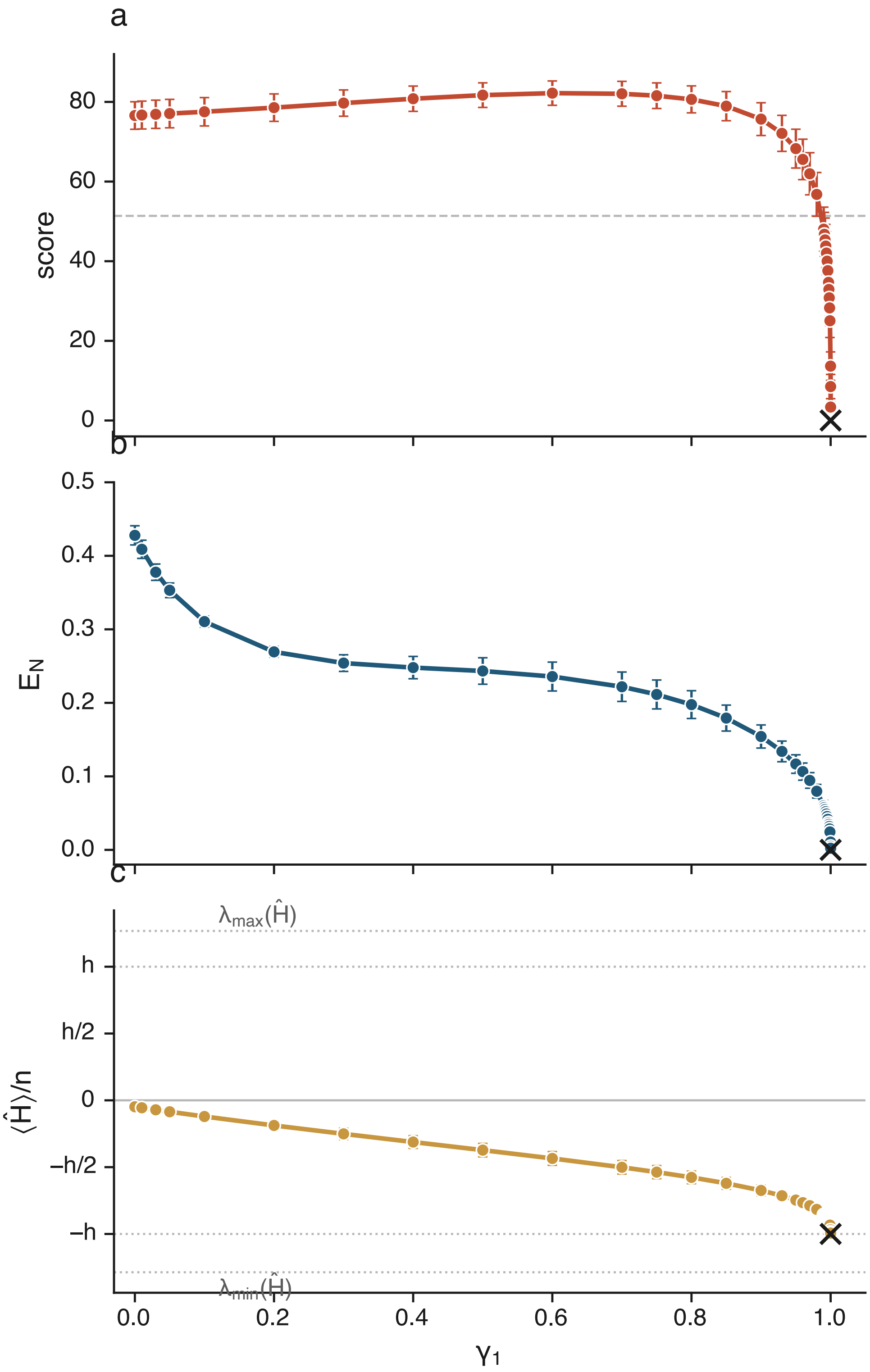


FIG. S8. **Moderate relaxation raises mean score while mean $E_N$ falls.** Same 25 survey cells and closed-system $J^{\star}(n, N_M)$ as main-text Figs. 3 and 4, scored along the dissipation ladder (dephasing off). Equal-cell-weight means; bars are the standard error across cells. **(a)** Mean score. Dashed: linear baseline 51.4. Cross: complete damping ($\gamma_1 = 1$, score $= 0$). **(b)** Mean $E_N$. **(c)** Mean energy per qubit; complete damping reaches all $|0\rangle$.

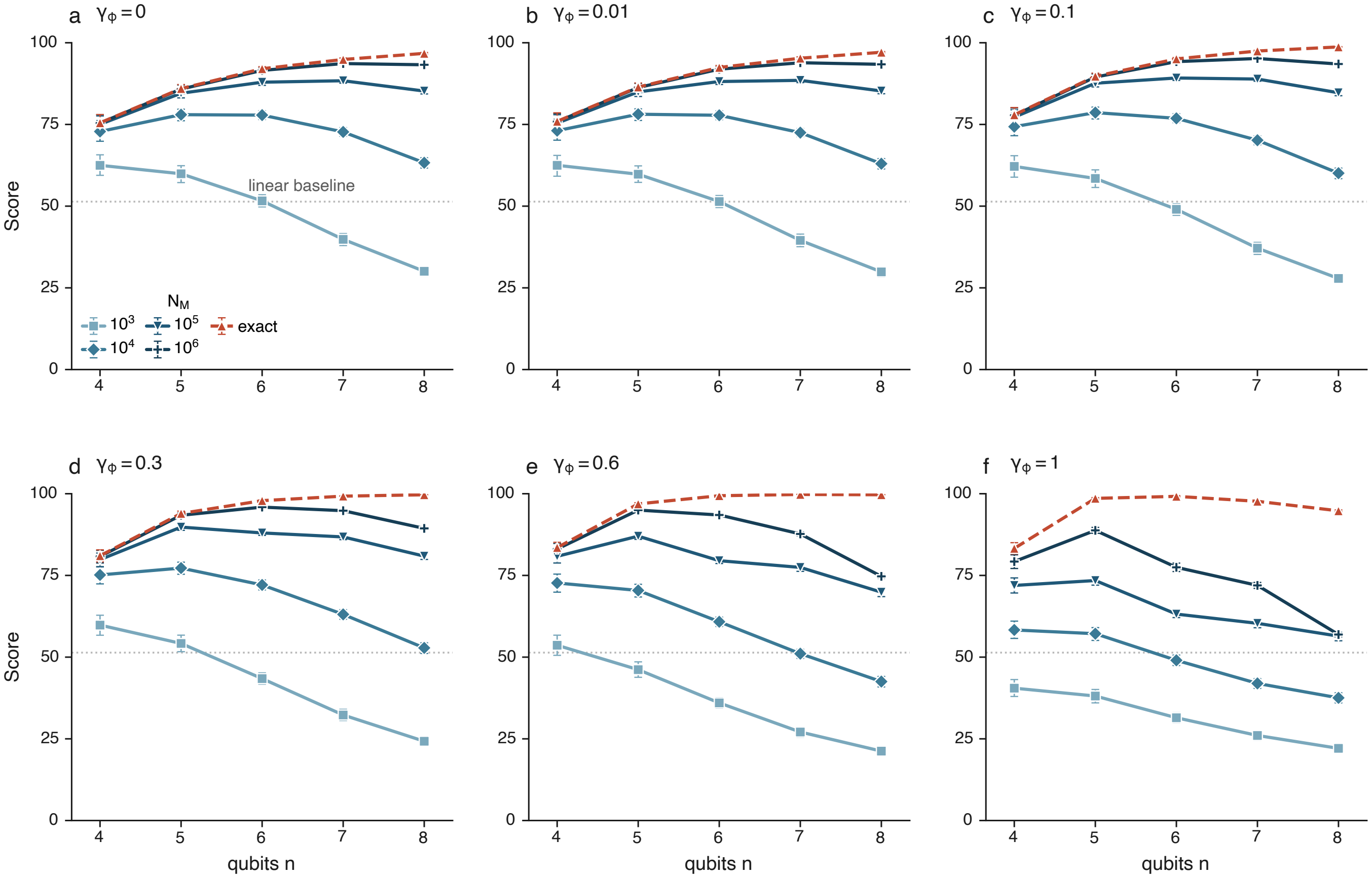


FIG. S9. **Score versus size at six dephasing strengths** ($\alpha = 0$). Same closed-system $J^\star(n, N_M)$ cells as main-text Fig. 3. **(a–f)** $\gamma_\varphi = 0, 0.01, 0.1, 0.3, 0.6, 1$. Curves are $N_M = 10^3$–$10^6$ and exact readout. Dotted line: linear baseline 51.4. Panel (a) is main-text Fig. 3a.

## XI. WASHOUT AND MACKEY–GLASS SANITY CHECKS

The main scores discard the first 100 steps. Figure S10a checks that this washout is long enough for the initial state to be forgotten. Two copies of one reservoir start in different pure states and then see the same inputs. By the last washout step their trace distance is at most $3.5 \times 10^{-10}$.

Figure S10b repeats the selection task with exact expectation values, once on i.i.d. streams and once on two independent Mackey–Glass series (delay $\tau = 30$). The ensemble is six reservoirs, two stream pairs, and one cue sequence, at two couplings on the main grid. Valid-cue scores stay above the shuffled-cue scores and above a linear readout of the raw inputs.

Figure S10c is the conventional one-series forecast, not the selection task. One Mackey–Glass series (delay $\tau = 17$) drives the reservoir, and the readout is trained on that same series $h$ steps ahead. Six reservoirs, one trajectory, exact expectation values. One step ahead, a linear readout of the current sample already scores 98.0. Further ahead that memoryless readout follows the autocorrelation of the series, while the reservoir stays high: at 32 steps the mean scores are 93.2 ($J_s/h = 0.32$) and 89.6 ($J_s/h = 1.58$). Figure S11 shows the target and the prediction for one example of each Mackey–Glass check. These checks are not the main performance map.

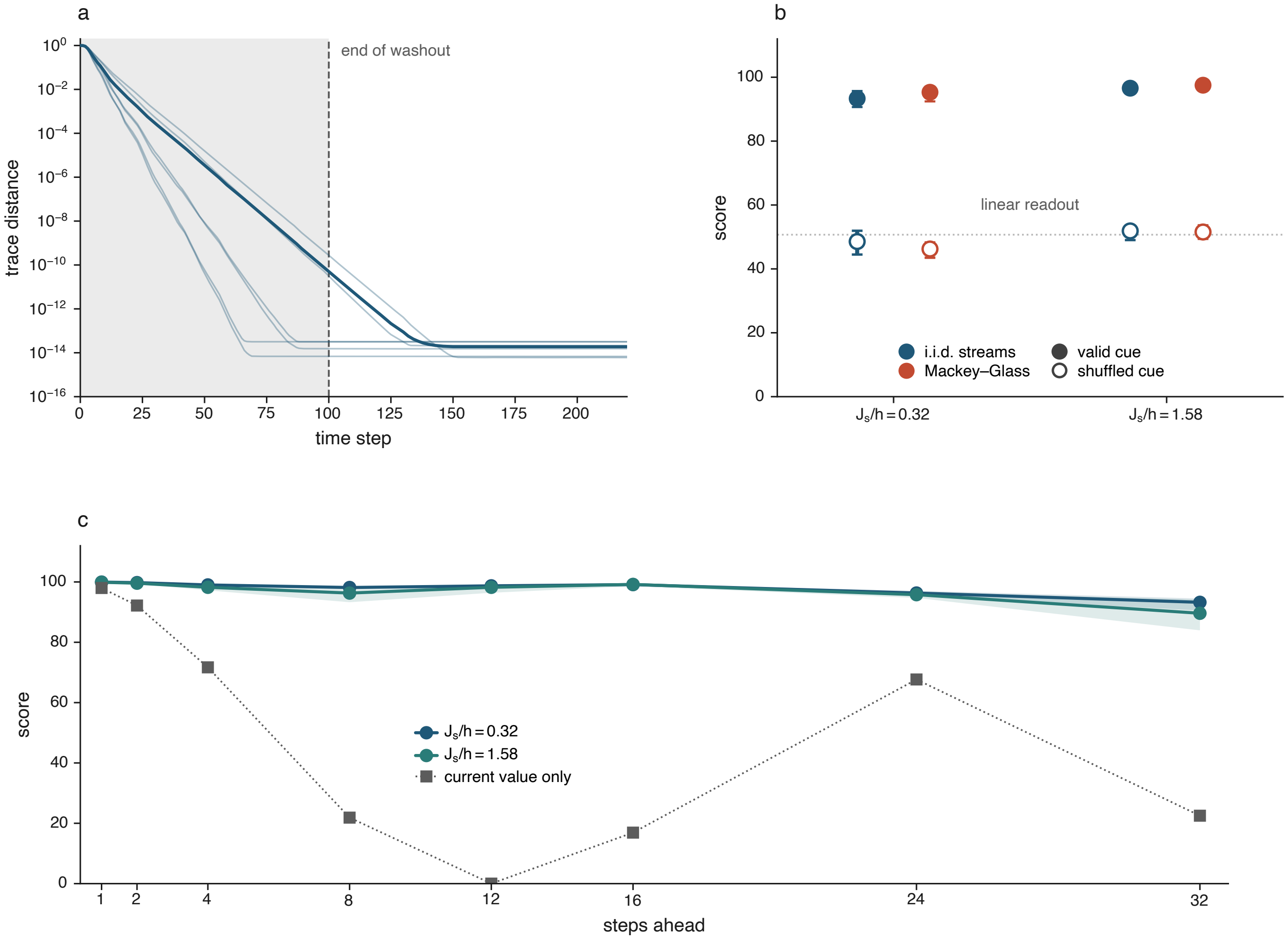


FIG. S10. **Sanity checks: washout, two-stream selection, and a one-series Mackey–Glass forecast.** Not the main performance map. **(a)** Trace distance $D = \frac{1}{2}\|\rho_A - \rho_B\|_1$ between two copies of one $n = 8$ reservoir ($J_s/h = 0.32$, no dephasing or relaxation). One copy starts in $|0\rangle^{\otimes n}$ and the other in $|1\rangle^{\otimes n}$; both then see the same inputs (one i.i.d. stream pair, one cue, $\alpha = 0$). Thin lines: six reservoirs. Solid line: their mean. The grey band is the washout (first 100 steps). At the last washout step, $D \leq 3.5 \times 10^{-10}$. By step 200 every copy is at numerical roundoff ($\sim 10^{-14}$). **(b)** Same selection task, exact expectation values. Navy: i.i.d. uniform streams. Rust: two independent Mackey–Glass series (delay $\tau = 30$), each standardized. Filled: valid cue. Open: shuffled cue. Points are means over six reservoirs, two stream pairs, and one cue; bars are the 2.5–97.5 percentiles across reservoirs. Dotted line: linear readout of the raw inputs on the main i.i.d. atlas (50.7). At $J_s/h = 0.32$, valid-cue scores are 93.3 (i.i.d.) and 95.3 (Mackey–Glass); shuffled-cue scores are 48.6 and 46.2. At $J_s/h = 1.58$, valid-cue scores are 96.6 and 97.5; shuffled-cue scores are 51.9 and 51.5. Valid-cue scores exceed both the shuffled-cue scores and the linear readout. **(c)** Conventional one-series forecast. One Mackey–Glass series, delay $\tau = 17$, min-max scaled to $[-1, 1]$, injected with the same reset map. Teacher-forced: the reservoir sees $s(t)$ and the target is $s(t + h)$. Exact expectation values. Navy: $J_s/h = 0.32$. Teal: $J_s/h = 1.58$. Bands: 2.5–97.5 percentiles over six reservoirs. Dotted: linear readout of the current sample only, same score. One step ahead that readout scores 98.0; the reservoir scores 100 and 99.9. At 32 steps the reservoir means are 93.2 and 89.6; the current-sample readout scores 22.5. The dotted curve is not monotone because the series oscillates: the current sample is almost uncorrelated with the value 12 steps ahead (score 0) and correlated again 24 steps ahead (score 67.7).

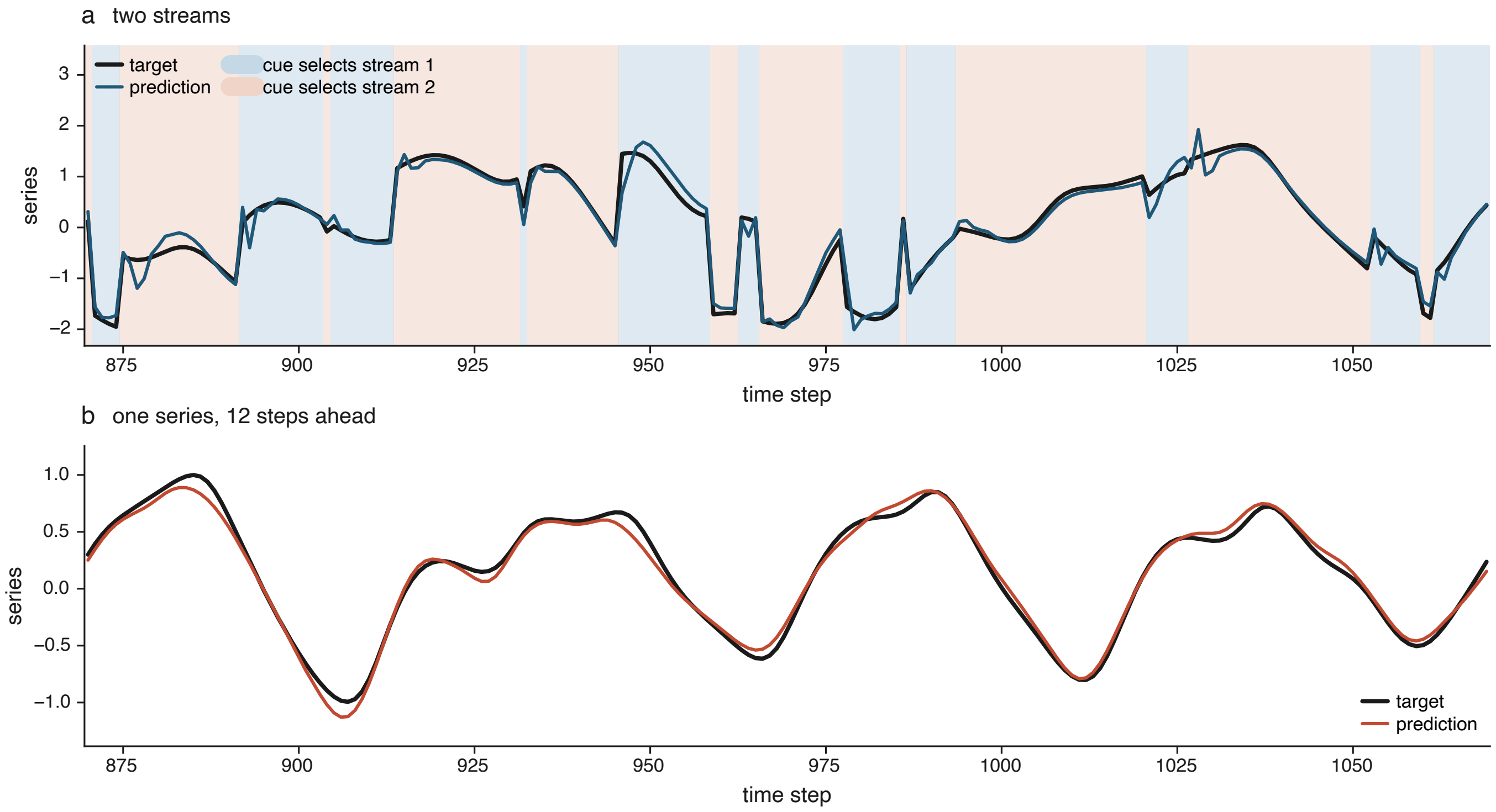


FIG. S11. **Target and prediction for the two Mackey–Glass checks.** One reservoir (seed 0, $J_s/h = 0.32$), exact expectation values. The curves are the first 200 steps of the test window. The quoted score is for the whole test window of this example, not the ensemble mean in Fig. S10. **(a)** Two-stream selection, delay $\tau = 30$, stream pair 0, cue sequence 0, valid cue, $\alpha = 0$. Black: target. Navy: prediction. This example scores 97.0. Blue band: the cue selects stream 1. Peach band: stream 2. **(b)** One-series forecast, delay $\tau = 17$, target 12 steps ahead. Black: target. Rust: prediction. This example scores 98.9. A linear readout of the current sample scores 0 at this horizon.

## XII. ADDITIONAL METHODS

These notes document how the operational success criterion was defined and how the entanglement–score association was estimated. In the main text, reservoirs in the same condition tend to score higher when their entanglement is higher (Fig. 4). That statement summarizes a within-cell Spearman rank correlation of score with $E_N$, averaged over cells. The correlation values are not given in the main text. The SI figures that show a fitted slope use a related within-condition regression.

### A. Stream cross-talk

The atlas cube and several SI controls mix the two streams before injection. The mixed pre-inputs and the coefficient $\alpha$ are Eq. (S1) in Sec. I. The normalization keeps the marginal variance of $v_1$ and $v_2$ identical for every $\alpha$, so increasing $\alpha$ increases stream overlap without leaking the cue through input amplitude. The atlas uses $\alpha \in \{0, 0.3, 0.5, 0.7\}$; $\alpha = 1$ is singular and excluded. The main-text figures use unmixed streams ($\alpha = 0$); the other values are Fig. S1. When mixing is on, the reservoir is driven by $(v_1, v_2)$ and the target remains $y_t = c_t u_{1,t} + (1 - c_t) u_{2,t}$.

### B. Score and baselines

Score is as in the main text. Valid-cue and shuffled-cue drives are scored separately. The shuffled drive uses the same reservoir and streams with a block-preserving circular shift of the cue, so the cue is present but uninformative.

Two input-only references use the same splits and a tuned lag $K \in \{1, 2, 4, 8\}$. A linear readout of lagged $(v_1, v_2, c)$ scores 50.7 on the eight-qubit atlas and 51.4 on the size survey, and cannot form the cue–stream product. A bilinear readout that is also given the products of the cue with the two streams, $(c\,v_1, c\,v_2)$, scores 100, the perfect score in the main text.

### C. Operational success criterion

The main text calls a setting successful when the valid-cue score exceeds the linear baseline and exceeds the shuffled-cue score. The operating region drawn on the SI maps uses those two comparisons and a third numerical filter: the valid–shuffled gap must be at least 0.25 of the gap from the shuffled score to the bilinear reference (0.50 for a stricter cut). The filter uses only the scores. 92 settings pass; 40 of them have a finite number of measurements.

### D. Within-condition ranking of score by entanglement

Main-text Fig. 4 compares reservoirs inside each $(n, N_M)$ cell, and again inside each $(n, N_M, \gamma_\varphi)$ cell, at the coupling $J^\star$ that maximizes mean score at $\gamma_\varphi = 0$. Within the cell, $E_N$ and score are demeaned and the grand mean is restored. The Spearman rank correlation of those values is then averaged over cells. The SI slope panels (Supplementary Figs. S6 and S7) fit a related atlas regression: within each successful setting at a finite number of measurements, $E_N$ is standardized across reservoirs, and score (or valid minus shuffled) is regressed on that standardized $E_N$. The slope is the change in score points associated with a one-standard-deviation increase in $E_N$ at fixed operating conditions. It is not a coefficient of the trained readout.

The raw fit uses only the setting intercepts. A feature-quality adjustment additionally controls for a bilinear interaction probe and the task-aligned signal-to-noise of the features. That adjustment reduces the slope but does not remove it ($+0.47\,[+0.30, +0.65]$ versus raw $+1.23$; about 62% of the raw slope is absorbed). Mechanism controls (feature standard deviation, effective rank, connected $ZZ$, stream-decode imbalance) leave the slope essentially unchanged. Score itself is never used as a control.

### E. Held-out reservoirs

Supplementary Figs. S6 and S7 report this comparison, the split, and the result. A setting is one coupling $J_s/h$, one stream mixing $\alpha$, and one number of measurements $N_M$. It is successful when the scores pass the checks in Sec. XII C. The 200 reservoirs are split in half, $r < 100$ and $r \geq 100$. One half decides which settings are successful.

The slope is fit only on the other half. The two halves then swap roles, and the two slopes are averaged. Inside such a setting, reservoirs with more entanglement have a higher score. In Fig. S6 most of that higher score is present for a scrambled cue as well as for a valid cue. In Fig. S7 the gap between those two scores is also larger, and a small positive remainder of that gap survives the feature-quality adjustment.

The uncertainty interval redraws whole reservoirs (3000 draws), so every stream and cue repeat of a reservoir stays together [2]. That interval leaves out some of the uncertainty in how the successful settings were chosen. Across both halves, 39 settings with a finite number of measurements are successful.

### F. Valid versus shuffled cue

Supplementary Fig. S6 repeats that fit on the valid-cue score and on the shuffled-cue score separately. Those two drives are the cue arms: the valid cue, and the same cue with its timing scrambled. The fit stays inside one operating setting, a shared coupling, stream mixing, and number of measurements. Much of the entanglement–score association is shared by the two arms; a smaller difference between the two slopes is the valid-minus-shuffled association. $E_N$ is measured on the valid-cue drive. A check that recomputes $E_N$ on the shuffled drive (40 reservoirs at two couplings) finds the two negativities nearly identical (within-setting Pearson $r = 0.996$), so the decomposition is not an artefact of which drive supplies $E_N$. This split is exploratory.

Binned SI panels remove each setting's mean from score and from $E_N$, bin the residual $E_N$, and add the global means back so the axes stay in physical units [3]. Those displayed lines are descriptive. They are separate from the within-cell association in main-text Fig. 4.

### G. Entanglement quantity

The recorded marker is the three-qubit input-block-versus-rest logarithmic negativity [4, 5] $E_N^{012|\overline{012}}$, the negativity across the $\{0, 1, 2\}$-versus-rest cut of the post-drive valid-cue state, averaged over snapshots after washout. On the eight-qubit atlas that average uses eight snapshots. The coherence in the main-text dephasing survey uses four snapshots. The block $\{0, 1, 2\}$ is the predeclared three-qubit input-containing partition, fixed before marker estimation—only qubit 0 is directly reset by the inputs, and qubits $1, 2$ complete the block without being spatial neighbors (the all-to-all model defines no spatial graph). There is no distinct cue qubit, since all three scalars are routed through qubit 0; the notation $E_N^{012|\overline{012}}$ is used to avoid any reading as a single-qubit-versus-rest cut. Other three-qubit input-containing cuts are compared in Supplementary Fig. S5. Entanglement is read out on the valid-cue drive only, is independent of $N_M$ (a state property), and is never used in region selection [6].

### H. Confirmatory versus exploratory analyses

The association stated in the main text is the within-condition comparison of score with the predeclared $\{0, 1, 2\}$ input-block cut on the size survey (main-text Fig. 4), using the Spearman correlation defined above. That fitted slope, the feature-quality adjustment, the cut sweep, and the cue-arm split (Sec. XII F) are secondary robustness and exploratory analyses: their intervals are presented descriptively and are *not* multiplicity-adjusted, so individual entries of those sweeps should not be read as independently confirmed findings. Cross-fitting limits direct seed-reuse bias from region selection but does not propagate all operating-region selection uncertainty. It also does not correct for testing many cuts, quantities, or dissipation strengths. The ensemble expansions (main performance map $50 \to 200$ reservoirs; scaling $24 \to 100$ master seeds) were preplanned power upgrades of the same predefined designs, executed by extending the seed ranges without any other change. The atlas slope moved from $+0.49\,[+0.20, +0.78]$ at the original power to $+0.47\,[+0.30, +0.65]$ at full power, i.e. the interval tightened around an unchanged central value rather than an effect being chased into significance.

### I. Dedicated simulation of the operating region

The cut sweep repeats the valid-cue drive at the 29 unique $(J, \alpha)$ pairs of the operating region at a finite number of measurements, for the original 50 reservoirs ($r = 0, \ldots, 49$) and the full stream×cue grid, with the same inputs as the main performance-map drive. It is not the later 200-reservoir ensemble. At 32 post-washout snapshots per drive it records the block negativity $E_N^{012|\overline{012}}$, the input-qubit negativity $E_N^{0|\bar{0}}$, the negativity of every three-qubit cut

containing qubit 0 (21 cuts), the mutual information $I(012\text{:rest}) = S(\hat{\rho}_A) + S(\hat{\rho}_B) - S(\hat{\rho})$, and the $\ell_1$ coherences of the full and reduced-012 states. Each quantity is then passed through the identical cross-fitted marker regression (state quantities are $N_M$-independent, so one drive covers every $N_M$ value at that pair).

### J. System-size and dissipation numerical study

Main-text Figs. 3 and 4 and Supplementary Figs. S8 and S9 come from a separate numerical study over sizes $n = 4, \ldots, 8$ with unmixed streams ($\alpha = 0$). The within-step dynamics, injection order, feature map, splits, and ridge selection follow the $n = 8$ map. This survey uses a different set of inputs. A linear readout of those inputs scores 51.4; the eight-qubit atlas baseline is 50.7. The tested numbers of measurements are $N_M \in \{10^3, 10^4, 10^5, 10^6, \infty\}$. For each of 100 master seeds, one random eight-qubit coupling matrix is drawn and reused at smaller sizes by keeping the first $n$ qubits, so the sizes share the same couplings and the same input qubit. The coupling value $J_s/h$ is taken from the same 13-point grid as the eight-qubit map ($\approx 0.20$–$3.16$). Each cell averages 4 stream pairs $\times\, 4$ cue sequences on the valid and shuffled arms. For each pair $(n, N_M)$, $J^\star(n, N_M)$ is the coupling that maximizes the mean score at $\gamma_\varphi = 0$. That coupling is held fixed as dephasing or relaxation increases.

A $T_1$-type channel toward $|0\rangle$ occupies the same slot as main-text dephasing (after $\mathcal{U}_\tau$, once per substep), with dephasing off. For one qubit with per-substep probability $p_1$ the Kraus operators are

$$\hat{K}_0 = |0\rangle\langle 0| + \sqrt{1 - p_1}\, |1\rangle\langle 1|, \qquad \hat{K}_1 = \sqrt{p_1}\, |0\rangle\langle 1|, \tag{S3}$$

with the $n$-qubit map the tensor product of the local channels. Supplementary Fig. S8 places the closed-system $J^\star(n, N_M)$ cells on an 36-point ladder on $[0, 1]$, applied after each of the three substeps with per-substep probability $p_1 = 1 - (1 - \gamma_1)^{1/3}$ (nominal cumulative parameter $\gamma_1$). Those conditions share the closed-system reservoirs and inputs. The dephasing sweep reuses that ladder as $\gamma_\varphi$ on the same cells, with relaxation off, and is the sweep reported in the main text. After each unitary substep the state is passed through independent local phase-flip channels of probability $p_\varphi = \frac{1}{2}\left[1 - (1 - \gamma_\varphi)^{1/3}\right]$. The cube root has the same role as in the main text. The channel follows each of the three injections, and $\gamma_\varphi$ is the suppression over that whole logical step, so three applications multiply a coherence of Hamming distance $d_H$ by $(1 - \gamma_\varphi)^{d_H}$. The same division across the three substeps makes $p_1$ compound to $\gamma_1$. Coherences in the dephasing sweep pick up a Hamming factor $(1 - 2p_\varphi)^{d_H}$ at each substep. When the nominal parameters are equal, $\gamma_\varphi = \gamma_1$, that factor equals $(1 - p_1)^{d_H}$. Each sweep uses one channel. At $\gamma_\varphi = 1$ the state is diagonal in the computational basis and is not forced to $|0 \cdots 0\rangle$.

Main-text Figs. 3 and 4 and Supplementary Figs. S8 and S9 report the unnormalized logarithmic negativity $E_N$ of qubits $\{0, 1, 2\}$ versus the rest. For this cut $E_N$ is at most $\min(3, n - 3)$ ebits. The within-cell association in main-text Fig. 4 uses the Spearman correlation of score with that $E_N$, at the fixed coupling $J^\star(n, N_M)$. The ratio $E_N / \min(3, n - 3)$ is the three-qubit predictor in the mixed-stream size scan of Supplementary Fig. S7c.


\advance\sipage by 1
\ifnum\sipage>\pdflastximagepages\else\repeat

\end{document}